\documentclass[10pt]{iopart}
\usepackage[dvips]{graphicx}
\usepackage{dcolumn}
\usepackage{bm}
\usepackage{dsfont}
\usepackage{color}
\usepackage{url}
\usepackage[colorlinks=true,citecolor=blue]{hyperref}

\usepackage{iopams}

\def\bs#1{\boldsymbol{#1}}
\def\txt#1{\textnormal{#1}}

\def\be{\begin{equation}}
\def\ee{\end{equation}}

\newcommand{\rec}{p}           
\newcommand{\brec}{\bs p}      

\begin{document}

\title[Measuring topology in a laser-coupled honeycomb lattice]{Measuring topology in a laser-coupled honeycomb lattice: \\ From Chern insulators to topological semi-metals}

\author{N. Goldman}
\ead{ngoldman@ulb.ac.be}
\address{Center for Nonlinear Phenomena and Complex Systems,\\
Universit\'{e} Libre de Bruxelles (U.L.B.), B-1050 Brussels, Belgium}
\author{E. Anisimovas}
\ead{egidijus.anisimovas@ff.vu.lt}
\address{Department of Theoretical Physics, Vilnius University, 
Saul\.{e}tekio 9, LT-10222 Vilnius, Lithuania}
\address{Institute of Theoretical Physics and Astronomy, Vilnius University, \\
A.~Go\v{s}tauto 12, LT-01108 Vilnius, Lithuania}
\author{F. Gerbier}
\address{Laboratoire Kastler Brossel, CNRS, ENS, UPMC, 24 rue Lhomond, 75005 Paris}
\author{P. \"{O}hberg}
\address{SUPA, Institute of Photonics and Quantum Sciences, Heriot-Watt University, EH14 4AS, Edinburgh, United Kingdom}
\author{I. B. Spielman}
\address{Joint Quantum Institute, National Institute of Standards and Technology,\\
and University of Maryland, Gaithersburg, Maryland, 20899, USA}
\author{G. Juzeli\={u}nas}
\ead{gediminas.juzeliunas@tfai.vu.lt}
\address{Institute of Theoretical Physics and Astronomy, Vilnius University, \\
A.~Go\v{s}tauto 12, LT-01108 Vilnius, Lithuania}

\date{\today}


\newpage

\begin{abstract} 
Ultracold fermions trapped in a honeycomb optical lattice constitute a versatile setup to experimentally realize the Haldane model [\emph{Phys. Rev. Lett. {\bf 61}, 2015 (1988)}]. In this system, a non-uniform synthetic magnetic flux can be engineered through laser-induced methods, explicitly breaking time-reversal symmetry. This potentially opens a bulk gap in the energy spectrum, which is associated with a non-trivial topological order, i.e., a non-zero Chern number. In this work, we consider the possibility of producing and identifying such a robust Chern insulator in the laser-coupled honeycomb lattice.  We explore a large parameter space spanned by experimentally controllable parameters and obtain a variety of phase diagrams, clearly identifying the accessible topologically non-trivial regimes. We discuss the signatures of  Chern insulators in cold-atom systems, considering available detection methods. We also highlight the existence of topological semi-metals in this system, which are gapless phases characterized by non-zero winding numbers, not present in Haldane's original model.

\end{abstract}

\maketitle

\section{Introduction}
Topological phases of matter have been a topic of great interest in condensed matter physics since the discovery of the integer quantum Hall effect  \cite{vonKlitzing:1986}. They are characterized by transport properties -- such as a quantized Hall conductivity -- that depend on the topological structure of the eigenstates \cite{Kohmoto:1985}, and not on the details of the microscopic Hamiltonian. As a result, such properties are remarkably robust  against external perturbations. Integer quantum Hall phases, the first topological insulating phases to be discovered \cite{vonKlitzing:1986}, are realized by applying a large uniform magnetic field to a quasi-ideal two-dimensional electron gas, as formed in layered semiconductors structures. 

The presence of a uniform magnetic field is not, however, a necessary condition to produce quantum Hall states, as first realized by Haldane \cite{Haldane1988}. He proposed
a remarkably simple model on a honeycomb lattice, with real nearest-neighbor (NN) hopping and \emph{complex} next-nearest neighbor (NNN) hopping mimicking the Peierls phases experienced by charged particles in a magnetic field. Although the magnetic flux through an elementary cell of the honeycomb lattice is zero, a staggered magnetic field present within this cell locally breaks time-reversal symmetry. Haldane showed that this model supports phases that are equivalent to integer quantum Hall phases: they correspond to insulators with quantized Hall conductivities, $\sigma_H=\nu \, e^2/h$ where $e$ is the electron charge. In this manner, it is possible to generate a quantum Hall effect without a uniform external magnetic field. The integer $\nu =\pm1$ (depending on the particular values of the microscopic parameters) is a topological invariant  -- the Chern number -- characteristic of the phase and robust with respect to small perturbations \cite{Thouless1982,Kohmoto:1985}. More recently, a more broad concept of  topological insulators has emerged, classifying all possible topological phases for non-interacting fermions in terms of their symmetries \cite{Hasan2010,Qi2011}. In this modern terminology, the Haldane model belongs to the class A of Chern insulators, which are topologically equivalent to the standard quantum Hall states.

The Haldane model has not been directly realized in solid-state systems, due to the somewhat artificial structure of the staggered magnetic field. Interestingly, ultracold atomic gases \cite{Lewenstein:2007,Bloch2008a} appear better suited to achieve this goal \cite{Liu:2010,Stanescu:2010}. In recent years, many proposals have been put forward to realize artificial magnetic fields for ultracold atoms (see \cite{Dalibard2011} for a review). Staggered fields are relatively easier to implement than uniform ones \cite{Jaksch2003,Gerbier2010,Lim2008}, and have already been realized in a square optical lattice \cite{aidelsburger2011}. Building on these ideas, Alba and coworkers \cite{Alba2011} proposed a model very similar to Haldane's that could be realized with ultracold atoms. Their variant is based upon a state-dependent honeycomb optical lattice \cite{Bloch2008a}, where cold atoms in two different internal ``pseudospin'' states are localized at two inequivalent sites of the elementary cell. Additionally, laser induced transitions \cite{Jaksch2003,Ruostekoski:2002} between the nearest-neighbor sites lead to pseudospin-dependent hopping matrix elements containing phase factors, schematically depicted in Fig.~\ref{figlattice}. Furthermore, Alba and coworkers \cite{Alba2011} suggested a measurement based on spin-resolved time-of-flight (ToF) experiments to identify topological invariants.  \\

The present work provides a systematic analysis of the model proposed in Ref. \cite{Alba2011} and identifies parameter regimes where Chern insulators emerge. The goals are: firstly, to serve as a detailed guide to possible experiments aiming at realizing such topological phases; and secondly, to discuss the subtle issue of identifying them through ToF methods. Ref. ~\cite{Alba2011} focused on the very special case  when one of the NNN hopping amplitudes was zero, and we find that such a system is semi-metallic, not a quantized Chern insulator. In this limit, where the bulk energy gap is \emph{closed}, the Hall conductivity is no longer simply given by a Chern number $\sigma_H \ne \nu \, e^2/h$, and therefore, this transport coefficient generally looses its topological stability. Yet, the ToF method seems to give a non-trivial signature in this regime, which is robust with respect to small variations of model parameters. The subtlety is that the ToF method of Ref. \cite{Alba2011} actually measures a winding number \cite{Qi:2006}, which only coincides with a topologically protected Hall conductivity \emph{when the energy gap is open}. If this condition is met, the ToF method of Ref. \cite{Alba2011} then produces a reasonable experimental measure of the topologically invariant Chern number, and we indeed verify its robustness when varying the system parameters. Interestingly, if the bulk gap is closed, we find that the winding number measured from a ToF absorption image might still depict a stable plateau when varying the microscopic parameters, under the condition that the Fermi energy is exactly tuned at the gap closing point. In this work, such gapless phases associated with a non-trivial winding number will be referred to as \emph{topological semi-metals}. Absent in the original Haldane model, they constitute intriguing topological phases, which can be created and detected in the laser-coupled honeycomb lattice. 

In this work, several types of band structures and topological orders will therefore be present: (1) Chern insulating phases, i.e. gapped phases with non-trivial Chern numbers $\nu=\pm 1$, (2) Topological semi-metals, i.e. gapless phases associated with a non-trivial winding number, (3) Standard semi-metals, i.e. gapless phases with the two bands touching at the Dirac points, as in graphene \cite{Wallace1947,CastroNeto2009}, and which are found at the transition between two topological phases. \\

This paper is structured as follows. In Sect. \ref{modelsection}, we introduce the model and discuss how the energy band topology can be characterized in terms of Chern numbers. We also discuss the magnetic flux configuration as a function of the model parameters, highlighting the time-reversal-symmetry breaking regimes. Sect. \ref{topologicalphases} presents the main results, where the phase diagrams are investigated as a function of the microscopic parameters. In Sect. \ref{skyrmion}, we examine the signatures of the ToF method \cite{Alba2011}, and compare its results when applied to a Chern insulator or to a semi-metallic phase, i.e., when the topological bulk gap is absent. We summarize the results in Sect. \ref{conclusions}, and discuss an extension which implements the Kane-Mele model leading to  $\mathbb{Z}_2$ topological insulators \cite{Kane:2005}.

\section{The Model and the gauge structure}
\label{modelsection}

\subsection{The Hamiltonian}

\begin{figure}
\centering
\includegraphics[width=1\columnwidth]{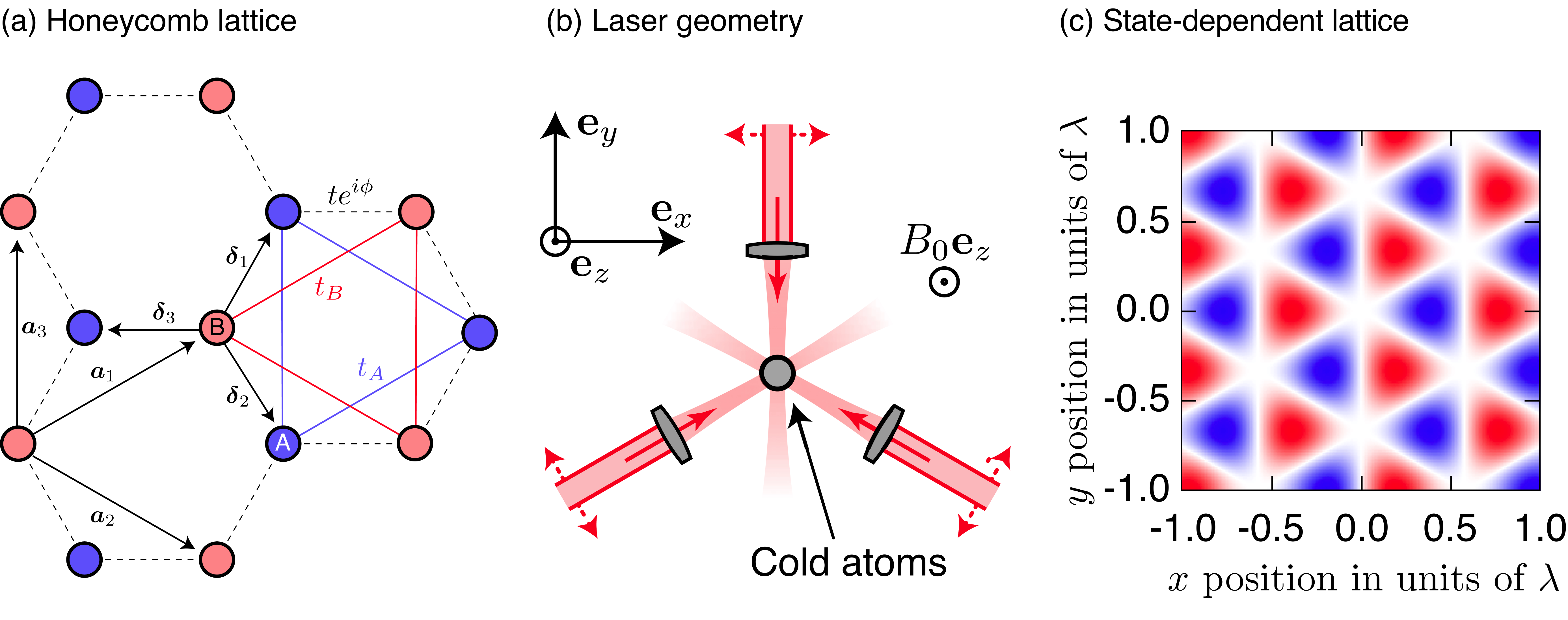}
\caption{\label{figlattice} (a) Honeycomb lattice composed of two coupled triangular sublattices $A$ and $B$. The site positions in each sublattice are defined as $\bs{r}_{m_A}=m_1 \bs a_1 +m_2 \bs a_2 $ and $\bs{r}_{m_B}=m_1 \bs a_1 +m_2 \bs a_2 -\bs \delta_2$, with unit vectors  $\bs a_1 = \bs \delta_1-\bs \delta_3$ and $\bs a_2 = \bs \delta_2-\bs \delta_3$ and with $m=(m_1,m_2)$ integer. The nearest-neighbour vectors are $\bs \delta_1= a/2 (1, \sqrt{3})$, $\bs \delta_2= a/2 (1, -\sqrt{3})$ and $\bs \delta_3= a(-1,0)$. We define $\bs a_3 = \bs a_1- \bs a_2= \bs \delta_1-\bs \delta_2$. The hopping factors between NN and NNN sites of the honeycomb lattice are indicated by $t_A$, $t_B$ and $t e^{i \phi}$, with $\phi\equiv \phi(m_A,m_B)$ given by Eq. \eqref{peierls}. The lattice spacing is $a\sqrt{3}$, and we set $a=1$ in the main text, which defines our unit of length. (b) Three-beam laser configuration giving rise to the desired spin-dependent hexagonal lattice, which we describe for $^{40}K$.  In this vision, the lasers are detuned between the D1 and D2 lines of the 4S-4P transition, whereby the state-independent (scalar) light shift is zero.  The remaining spin-dependent potential -- an effective Zeeman magnetic field -- is depicted in (c).  The strength of the ``same-spin" hopping, i.e. $t_a$ and $t_b$, is governed by the choice of internal states: the pair $\left|f = 9/2, m_F= 7/2\right>$ and $\left|f = 7/2, m_F= 7/2\right>$ produce $t_a\approx t_b$ as they have opposite magnetic moments. In contrast, the choice $\left|f = 9/2, m_F= 9/2\right>$ and $\left|f = 7/2, m_F= 7/2\right>$ produces $t_a\neq t_b$.  The effective Zeeman shift is plotted with a color scale where blue indicates the potential minima for pseudo-spin up atoms, forming the A sublattice; and red indicates the minima for pseudo-spin down atoms, forming the B sublattice. Not shown are an additional pair of Raman lasers, also in the ${\bf e}_x\!-\!{\bf e}_y$ plane, that couple between the different sublattices (red and blue in (c)). } 
\end{figure}

In the model introduced in Ref. \cite{Alba2011}, cold fermionic atoms are trapped in a honeycomb structure formed by two intertwined triangular optical lattices, whose sites are labeled by $A$ and $B$ respectively [cf. Fig. \ref{figlattice} (a)-(c)]. In the tight-binding regime -- applicable for sufficiently deep optical potentials $V_{A,B} (\bs x)$ -- atoms are only allowed to hop between neighboring sites of the two triangular sublattices, which correspond to next-nearest neighbors of the honeycomb lattice (denoted $\langle \langle n_{\tau}, m_{\tau} \rangle \rangle$, with $\tau=A,B$). The second-quantized Hamiltonian takes the form
\begin{equation}
\hat H_{\text{NNN}}= - t_A \sum_{\langle \langle n_A, m_A \rangle \rangle} \hat a^{\dagger}_{n_A} \hat a_{m_A} - t_B \sum_{\langle \langle n_B, m_B \rangle \rangle} \hat b^{\dagger}_{n_B} \hat b_{m_B},
\end{equation}
where $\hat a_{m_A}$ ($\hat b_{m_B}$) is the field operator for annihilation of an atom at the lattice site $\bs r_{m_A}$ ($\bs r_{m_B}$) associated with the $A$ ($B$) sublattice, and where $t_{A,B}$ are the tunneling amplitudes. Furthermore, the two sublattices are coupled through laser-assisted tunneling, where hopping is induced between neighboring sites of the honeycomb lattice by a laser coupling the two internal states associated with each sublattice. This corresponds to tunneling processes linking nearest neighbors sites of the honeycomb lattice, denoted as $ \langle m_{A}, m_{B} \rangle $, which are described by the Hamiltonian
\begin{equation}
\hat H_{\text{NN}}= - t \sum_{\langle m_A, m_B \rangle} \left( e^{i \phi (m_A, m_B)} \hat a^{\dagger}_{m_A} \hat b_{m_B} + h.\,c. \right).
\end{equation}
Here, the phases $\phi (m_A, m_B)$ generated by the laser fields are the analogs of the Peierls phases familiar from condensed matter physics \cite{Luttinger1951,Hofstadter1976}, with $\bs r_{m_A}$ and $\bs r_{m_B}$  specifying the nearest neighboring sites of the hexagonal lattice. Following the approach of Jaksch and Zoller \cite{Jaksch2003}, these phases can be expressed in terms of the momentum $\brec$ transferred by the laser-assisted tunneling as
\begin{equation}
\phi (m_A, m_B)= \brec \cdot (\bs r_{m_A} + \bs r_{m_B})/2 = - \phi (m_B,m_A),
\label{peierls}
\end{equation}
so that the phases have opposite signs for $\bs r_A \rightarrow \bs r_B$ and $\bs r_B \rightarrow \bs r_A$ hoppings (cf. Fig. \ref{figfluxlattice}(a) and Refs. \cite{Dalibard2011,Gerbier2010}). Finally, the model also features an on-site \emph{staggered} potential, described by
\begin{equation}
\hat H_{\text{stag}}= - \varepsilon \sum_{m} \left( \hat a^{\dagger}_{m_A} \hat a_{m_A} - \hat b^{\dagger}_{m_B} \hat b_{m_B}\right),
\end{equation}
which explicitly breaks the inversion symmetry of the honeycomb lattice \cite{Haldane1988}. The total Hamiltonian, given by 
\be
\hat H_{\text{tot}}=\hat H_{\text{NN}}+\hat H_{\text{NNN}}+\hat H_{\text{stag}},\label{htot}
\ee 
is characterized by the hopping amplitudes ($t$, $t_A$ and $t_B$), the momentum transfer $\brec =(p_x,p_y)$, as well as the mismatch energy $\varepsilon$. 

To eliminate the explicit spatial dependence of our Hamiltonian \eqref{htot}, we perform the unitary transformation
\begin{align}
&\hat a^{\dagger}_{m_A} \rightarrow \tilde a^{\dagger}_{m_A}=\hat a^{\dagger}_{m_A} \exp(i \brec \cdot \bs r_{m_A} /2 ), \\
&\hat b^{\dagger}_{m_B} \rightarrow \tilde b^{\dagger}_{m_B}=\hat b^{\dagger}_{m_B} \exp(- i \brec \cdot \bs r_{m_B} /2 ), \nonumber
\end{align}
giving a transformed Hamiltonian  
\begin{align}
\hat H_{\text{tot}}=& - t \sum_{\langle n_A, m_B \rangle} \bigl( \tilde a^{\dagger}_{n_A} \tilde b_{m_B} + \tilde b^{\dagger}_{m_B} \tilde a_{n_A} \bigr ) \nonumber \\
&- t_A \sum_{\langle \langle n_A, m_A \rangle \rangle} e^{i \tilde \phi (n_A, m_A)} \tilde a^{\dagger}_{n_A} \tilde a_{m_A} - t_B \sum_{\langle \langle n_B, m_B \rangle \rangle} e^{i \tilde \phi (n_B, m_B)} \tilde b^{\dagger}_{n_B} \tilde b_{m_B} \nonumber \\ 
& - \varepsilon \sum_{m}  \bigl( \tilde a^{\dagger}_{m_A} \tilde a_{m_A} - \tilde b^{\dagger}_{m_B} \tilde b_{m_B}  \bigr ), \label{hamnew}
\end{align}
with new Peierls phases given by
\begin{align}
&\tilde \phi (n_A, m_A)= \brec \cdot (\bs r_{m_A} - \bs r_{n_A})/2, \nonumber \\  
&\tilde \phi (n_B, m_B)= \brec \cdot (\bs r_{n_B} - \bs r_{m_B})/2.
\label{peierls2}
\end{align}
The transformed Hamiltonian \eqref{hamnew}, featuring complex hopping terms along the links connecting NNN sites, is similar to the Haldane model \cite{Haldane1988}, but with important differences highlighted in Sect. \ref{haldanesection}. 

Since $\bs r_{n_{A,B}} - \bs r_{m_{A,B}}=\bs \delta_{\mu}- \bs \delta_{\nu}=\pm \bs a_{\lambda}$ are the primitive lattice vectors of the honeycomb lattice (see Fig. \ref{figlattice}), where $\mu,\nu,\lambda=1,2,3$, the phases $\tilde{\phi}$ in Eq. \eqref{peierls2} no longer depend on the spatial coordinates. Therefore, the Hamiltonian \eqref{hamnew} is invariant under discrete translations, $[\hat H_{\text{tot}}, \mathcal{T}_{1,2}]=0$ where $\mathcal{T}_{1,2} \psi (\bs r)=\psi (\bs r+\bs a_{1,2})$,  allowing us to invoke Bloch's theorem and reduce the analysis to a  unit cell formed by two inequivalent sites $A$ and $B$. In momentum space, the Hamiltonian takes the form of a $2 \times 2$ matrix,
\begin{equation}
H(\bs k)= - \begin{pmatrix}
\varepsilon + 2 t_A f (\bs k - \brec / 2) & t g (\bs k) \\
 t g^{*} (\bs k) & - \varepsilon + 2 t_B f (\bs k + \brec / 2)
\end{pmatrix} ,
\end{equation}
where
\begin{align}
&f (\bs k)=\sum_{\nu=1}^{3}\cos \bigl ( \bs k \cdot \bs{a}_{\nu}  \bigr), \quad g (\bs k)=\sum_{\nu=1}^{3} \exp (-i \bs k \cdot \bs{\delta}_{\nu}), \label{gfunction}
\end{align}
and $\bs k =(k_x,k_y)$ belongs to the first Brillouin zone (FBZ) of the system. We rewrite this Hamiltonian in the standard form
\begin{equation}
  H(\bs k)= \epsilon (\bs k) \hat{1}+ \bs{d} (\bs k) \cdot \bs{\hat \sigma}, \label{hamtwo}
\end{equation}
with  
\begin{equation}
\epsilon (\bs k)= - t_A f (\bs k - \brec / 2) \label{epsilon_k}
    - t_B f (\bs k + \brec / 2),
\end{equation}
where $\bs {\hat \sigma}$ is the vector of Pauli matrices, and $\bs d (\bs k)$ has real-valued Cartesian components defined by
\begin{align}
    &d_x (\bs k)-id_y (\bs k)= - t g(\bs k), \nonumber \\
  &d_z (\bs k)= - \varepsilon - t_A f (\bs k - \brec / 2)
    + t_B f (\bs k + \brec / 2).\label{sz}
\end{align}
The eigen-energies of the Hamiltonian (\ref{hamtwo}) are $E_{\pm} (\bs k)=\epsilon (\bs k)\pm d (\bs k)$,  where we introduced the ``coupling strength" $d (\bs k)= \vert \bs d (\bs k) \vert$.  

Our Hamiltonian \eqref{hamtwo}-\eqref{sz} differs from the expression derived in Ref. \cite{Alba2011}, where different Peierls phases were used \footnote{In Ref. \cite{Alba2011}, Peierls phases were considered to be of the form $\phi (m_A, m_B)= \brec \cdot (\bs r_{m_A} - \bs r_{m_B})$ instead of Eq. \eqref{peierls}, cf. the Supplemental Material in Ref. \cite{Alba2011}. We note that the correct form \eqref{peierls}, used in the present work, corresponds to the synthetic Peierls phases that can be realized with cold atoms in optical lattices, following the method of Ref. \cite{Jaksch2003}.}. Both models are exactly equivalent when $t_{B} = 0$, where the system describes a semi-metal (cf.\ Section \ref{topologicalphases} and \ref{skyrmion}).

We now briefly describe how the energy spectrum changes with the parameters $(t_A,t_B, \varepsilon)$ of the Hamiltonian \eqref{hamtwo}-\eqref{sz}, in the absence of momentum transfer $\brec = 0$. When $t_{A,B}=\varepsilon=0$, the band structure is that of graphene \cite{Wallace1947,CastroNeto2009}, namely, the spectrum is given by
\begin{align}
E_{\pm} (\bs k)= \pm  \vert t  g(\bs k)\vert , \quad t_{A,B}=\varepsilon=0 .
\end{align} 
The two bands touch at zero energy for particular points $\bs K_{\pm}$ (the so-called Dirac points), where $g(\bs K_{\pm})=0$, and around which the spectrum is quasi-linear with momentum, $E_{\pm} (\bs k)\approx \pm v_F \vert \bs k \vert$. We will still use the term ``Dirac" points to denote $\bs K_\pm$, even if the gap is open (in the vicinity of these points the excitations describe massive Dirac fermions). For $\varepsilon \ne0$, a bulk gap $\Delta \propto \varepsilon$ opens at the Dirac points, where the gap width is defined as $\Delta = \text{min} (E_+) - \text{max} (E_-)$ \footnote{We set $\Delta =0$ when $ \text{max} (E_-) \ge  \text{min} (E_+)$. This happens when the two bands touch at a Dirac point, $E_+ (\bs K_{D}) = E_{-} (\bs K_{D})$, but also when the bulk gap is indirectly closed, cf. Figs. \ref{figa7} (a)-(c). The properties of semi-metallic phases with $\Delta=0$ are discussed in Section \ref{semimetal}.}. This is not a necessary condition to open a gap, as the NNN couplings $t_A,t_B$ are also able to do so. For $t_{A,B} \ne 0$ and $\varepsilon=0$, the spectrum is now 
\begin{align}
E_{\pm} (\bs k; \brec=0,\varepsilon=0)=D_{+}(\bs k) \pm \sqrt{ \vert t  ~g(\bs k)\vert]^2 +\left[ D_{-}(\bs k)  \right]^2 },
\end{align}
with $D_{\pm}(\bs k) = - (t_{A}\pm t_B) ~ f(\bs k)$. Next we note that $\vert g(\bs k) \vert^2 = 3  +2 f(\bs k)$, showing that a gap $\Delta \propto \vert t_A-t_B\vert$ opens at the Dirac points due to NNN couplings. For finite momentum transfer $\brec \ne 0$, the energy spectrum
\begin{align}
E_{\pm} (\bs k)= \epsilon (\bs k; \brec , t_{A,B}) \pm \sqrt{ \left[ t  ~g(\bs k)\right]^2 + \left[d_z (\bs k; \brec, \varepsilon , t_{A,B}) \right]^2 }
\end{align}
leads to more complex spectral structures and phases, to be explored in Section \ref{topologicalphases}. \\

In the following, we study the phases of non-interacting fermions in an optical-lattice setup described by Hamiltonian \eqref{hamtwo}-\eqref{sz}. Such a system forms a metal (or a semi-metal) when the gap is closed $\Delta =0$, and an insulator when $\Delta > 0$. In the latter case, we set the Fermi energy $E_{\text{F}}$ in the middle of the bulk gap. This classification in terms of the band structure is not exhaustive, and it must be completed by a description of the topological properties of this band structure. This is examined in the following Section \ref{chernsection}. In addition, the properties of some peculiar semi-metals are also explored in this work (cf. Section \ref{skyrmion}).
 
\subsection{The Chern number}\label{chernsection}

When the two-band spectrum $E (\bs k)$ exhibits an energy gap $\Delta$, one can define a topologically invariant Chern number  \cite{nakahara}, which encodes the topological order of the system. As shown in Ref. \cite{Thouless1982}, the Chern number $\nu$ is equal to the transverse Hall conductivity, $\sigma_H= \nu$ in units of the conductivity quantum, provided the \emph{Fermi energy is located in the bulk gap}. The Chern number is given by the standard TKNN expression \cite{Thouless1982,Xiao2010} 
\begin{align}
  \nu&=  \frac{i}{2 \pi} \int_{\mathbb{T}^2}
  \langle \partial_{k_x} u_{(-)}(\bs k) \vert \partial_{k_y} u_{(-)} (\bs k) \rangle
  - (k_x \leftrightarrow k_y) \txt{d}^2 \bs{k},\label{chern} \\
  &=  \frac{1}{2 \pi} \int_{\mathbb{T}^2} \bs 1_z \cdot (\nabla _{\bs k} \times \bs A (\bs k)  ) \txt{d}^2 \bs{k} ,\label{chernone}
\end{align}
where $\vert u_{(-)} (\bs k) \rangle$ denotes the single-particle eigenstate associated with the lowest bulk band $E_{-} (\bs k)$. The Berry's connection -- or vector potential -- $\bs{A}(\bs k)$ is defined by
\begin{align}
\bs A (\bs k) &= i \langle u_{(-)} \vert \bs \nabla _{\bs k} \vert u_{(-)} \rangle.\label{A-pm}
\end{align}
This quantity, which defines the parallel transport of the eigenstates over the FBZ \cite{nakahara}, also determines the  topological order of the system \cite{Kohmoto:1985}. The integration in Eq. \eqref{chernone} is taken over the FBZ, a two-torus denoted as $\mathbb{T}^2$, where the contribution due to any singularities of $\bs A(\bs k)$ -- to be discussed later on -- should be excluded.

It is convenient to parametrize the ``coupling" vector $\bs d (\bs k)$ in terms of the spherical angles $\theta \equiv \theta (\bs k)$ and $\phi \equiv \phi (\bs k)$, defined as
\begin{equation}
\tan \phi = d_y (\bs k) / d_x (\bs k) , \qquad \cos \theta= d_z (\bs k) / d (\bs k),\label{defangles}
\end{equation}
where $\phi= \pi- \arg g(\bs k)$ for $t>0$. In what follows we shall assume that $t>0$, without loss of generality. In this representation, the Hamiltonian \eqref{hamtwo} takes the form
\be
H (\bs k)= \epsilon (\bs k) \hat{1} - d (\bs k) \begin{pmatrix}
\cos \theta & e^{- i \phi} \sin \theta \\
e^{i \phi} \sin \theta & - \cos \theta
\end{pmatrix} . \label{hamcoupling}
\ee
The lowest eigenstate of \eqref{hamcoupling} is given by
\be
%
\qquad \vert u_{(-)} \rangle =\begin{pmatrix}
-e^{-i \phi} \sin (\theta /2) \\
\cos (\theta /2)
\end{pmatrix},\label{pm}
\ee
and from Eqs. (\ref{A-pm})-(\ref{pm}), we obtain an explicit expression for the Berry's connection,
\begin{equation}
\bs A (\bs k)=\frac{1}{2} (1- \cos \theta)   \bs \nabla _{\bs k} \phi . \label{berryconnection}
\end{equation}

A crucial point to note is that the Berry's connection \eqref{berryconnection} has singularities at the points in $\bs{k}$-space where $d_x (\bs k)= d_y (\bs k)=0$ and $d_z (\bs k)<0$. The condition $d_x (\bs k)= d_y (\bs k)=0$, which coincides with the zeros of the function $g (\bs k)$, is always fulfilled at the special points $\bs{K}_{\pm} = (\frac{2 \pi}{3}, \pm \frac{2 \pi}{3 \sqrt{3}})$,  where we have set $a = 1$. The second condition $d_z (\bs{K}_{\pm})<0$ is only satisfied for certain values of the model parameters ($t_{A,B}$, $\varepsilon$, $\brec$). In terms of the coupling vector $\bs d$, the singularity takes place at the ``South pole" where $\theta=\pi$ and $\phi$ is arbitrary, so that the state $\vert u_{(-)} \rangle$ is multivalued there.  Note that this singularity can be removed locally by a gauge transformation, but not globally \cite{Wu1975}. Moreover, we find that the phase $\phi= \pi- \arg g(\bs k)$ yields opposite vorticities at the two inequivalent Dirac points,
\be
v_{\pm}=\oint_{\gamma_{\pm}} \bs{\nabla} \phi (\bs k) \cdot \text{d} \bs k= \pm 2 \pi, \label{vortex}
\ee
where $\gamma_{\pm}$ denotes closed loops around the two Dirac points $\bs K_{\pm}$. 

If these singularities were absent, the integrand in Eq. \eqref{chernone} would constitute an \emph{exact} differential form over the entire FBZ. In this trivial case, Stokes theorem would then ensure that the integral in Eq. \eqref{chernone} is zero, since this exact two-form is integrated over a \emph{closed} manifold  \footnote{The FBZ is a two-dimensional torus $\mathbb{T}^2$, which is a closed manifold. See also \ref{analytical}.}. To account for these singularities, Stokes theorem can be applied to a contour avoiding them \cite{Kohmoto:1985,Hatsugai1993}. In particular, the Chern number \eqref{chernone} can be written as a sum of integrals performed over the excluded singularities, i.e. by contributions from small circles of infinitesimal radius $\gamma_{\pm}$ around the excluded Dirac points $\bs k=\bs K _{\pm}$ at which $\bs{A} (\bs k)$ is singular, 
\be
\nu = - \frac{1}{2 \pi} \sum_{\bs K _{-}, \bs K _{+}} \oint_{\gamma_{\pm}} \bs{A} (\bs k) \cdot \text{d} \bs k .
\ee 
Using Eq. \eqref{vortex} and taking into account the fact that $\cos \theta (\bs k)$ remains well-defined close to $\bs K _{\pm}$, we
find the simple expression for the Chern number
\begin{align}
  \nu&= \frac{1}{4 \pi} \left( \cos \left[ \theta \left(\bs K_{+}\right) \right] v_{+} ~+~ \cos \left[ \theta \left(\bs K_{-}\right) \right] v_{-} \right) \nonumber \\
  &=\frac{1}{2} \Biggl( \frac{d_z (\bs K_{+})}{\vert d_z (\bs K_{+}) \vert} - \frac{d_z (\bs K_-)}{\vert d_z (\bs K_{-}) \vert} \Biggr ), \label{chern2}
\end{align}
which only involves the sign of the ``mass" term $d_z (\bs k)$ \eqref{sz} at the two inequivalent Dirac points $\bs K_{\pm}$. A detailed demonstration of Eq. \eqref{chern2}, which further highlights the role played by the singularities, is presented in \ref{analytical}. The important result in Eq. \eqref{chern2} shows that the Chern number $\nu$ can now be directly evaluated, without performing the integration over the FBZ in Eq. \eqref{chern}. From Eqs. \eqref{sz}-\eqref{chern2}, one can already deduce that non-trivial Chern numbers $\nu \ne 0$ can only be obtained when $d_z (\bs k)$ has opposite signs at the two inequivalent Dirac points $\bs K_{\pm}$, which can only be achieved for $\brec \ne 0$. In the following Section \ref{fluxsection}, we give a physical interpretation in terms of effective magnetic fluxes and time-reversal-symmetry breaking. We also comment on pathological time-reversal symmetric configurations, which necessarily lead to a trivial topological order $\nu =0$.

To conclude this Section, we note that a Chern insulator is also characterized by current-carrying edge states that propagate along the edge of the system. This edge transport is guaranteed by the opening of a non-trivial bulk gap ($\Delta,\nu\ne0$, cf. Fig. \ref{figa7} (b)), and it leads to the quantization of the Hall conductivity {\it via} the bulk-edge correspondence \cite{Hatsugai1993}. The latter is observed through transport measurements in solid-state experiments.  In the cold-atom framework, such measurements are not convenient, as they would require atomic reservoirs coupled to the optical lattice. However, alternative methods, based on Bragg spectroscopy \cite{stenger1999b,steinhauer2002a}, have been proposed to extract and image these topological edge states \cite{Goldman2012}. We will use the appearance of chiral edge states later in this paper to strengthen the identification of Chern insulators (Section \ref{anisotropysection}). They are obtained from the spectrum of Hamiltonian (\ref{htot}) in a finite geometry \cite{Hatsugai1993}, as explained in the  \ref{app:edge}.

\subsection{Flux configurations and physical description of the model}
\label{fluxsection}

\begin{figure}
\centering
\includegraphics[width=1\columnwidth]{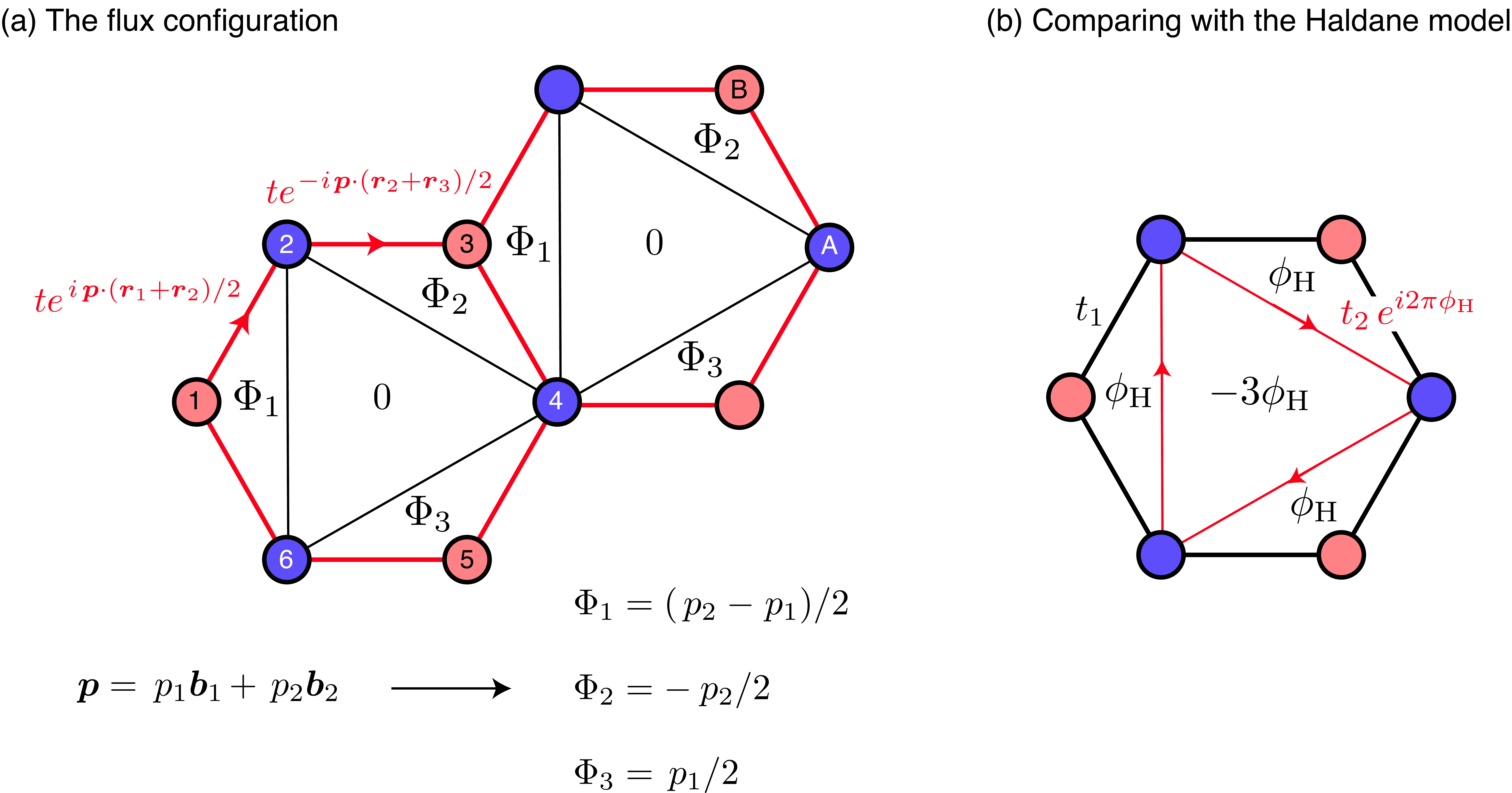}
\caption{\label{figfluxlattice} (a) Laser-coupled honeycomb lattice, including the Peierls phases \eqref{peierls}, and the corresponding flux configuration. The local fluxes $\Phi_{1,2,3}$ are explicitly given in terms of the momentum recoil $\bs p$. Here, the basic reciprocal lattice vectors are $\bs b_1=2 \pi/3 (1, \sqrt{3})$ and $\bs b_2=2 \pi/3 (1, -\sqrt{3})$. (b) The Haldane model and its simpler flux configuration, entirely characterized by the phase $\phi_{\text{H}}$.} 
\end{figure}

In this Section, we examine the effects of the Raman-induced phases in Eq. \eqref{peierls} from a less formal point of view, by associating  effective ``fluxes" to these Peierls phases. First, one can evaluate the number of magnetic flux quanta penetrating each hexagonal plaquette $\hexagon$, which yields (cf. Fig. \ref{figfluxlattice} (a))
\begin{align}
2 \pi \Phi(\hexagon) &= \sum_{\hexagon} \phi (n_A, m_B) \nonumber =0. 
\end{align}
Therefore, in the absence of NNN hopping (i.e. $t_{A,B}=0$), the system has a trivial flux configuration $\Phi=0$ and remains invariant under time reversal. 

Importantly, when NNN hopping terms are introduced (i.e. $t_{A,B} \ne 0$), triangular sub-plaquettes are penetrated by non-zero magnetic fluxes, explicitly breaking time-reversal symmetry and potentially leading to QH phases \cite{Haldane1988}. Considering the sub-plaquettes formed by the $A-B$ and $A-A$ hoppings, illustrated in Fig. \ref{figfluxlattice} (a), one finds that
\begin{align}
&\Phi_1=- \brec \cdot \bs a_3/4 \pi=(\rec_2 - \rec_1)/2 , \nonumber \\
 &\Phi_2=- \brec \cdot \bs a_2/4 \pi=-\rec_2/2, \nonumber \\
 &\Phi_3=\brec \cdot \bs a_1/4 \pi=\rec_1/2, \label{fluxeq}
\end{align}
where we expressed the recoil momentum $\brec=\rec_1 \bs b_1 + \rec_2 \bs b_2$ in terms of the basic reciprocal lattice vectors $\bs b_{1,2}$, for which $\bs b_{j}\cdot \bs a_{l}=2\pi \delta_{jl}$. The sub-plaquettes formed by the $A-B$ and $B-B$ hoppings have a similar flux structure. Thus, the space-dependent Peierls phases \eqref{peierls} produce a flux configuration characterized by three local fluxes $\Phi_{1,2,3}$, and which is translationally invariant over the whole lattice (cf. Fig. \ref{figfluxlattice} (a)). We also note that $\sum_{\alpha}\Phi_{\alpha}=0$, which indicates that the total flux penetrating each hexagonal plaquette $\hexagon$ remains zero, as found above \cite{Haldane1988}.

The system remains invariant under time reversal when $H(\{ \Phi_{1,2,3} \}) \equiv H (-\{ \Phi_{1,2,3} \})$, where $\{ \Phi_{1,2,3} \}$ represents the flux configuration stemming from a given $\brec$.  Besides the obvious case $\brec$=0, we find from Eq. \eqref{fluxeq}, that this occurs: 
\begin{itemize}
\item if $\rec_1$ and $\rec_2$ are \emph{both integers}, {\emph i.e.} if $\brec$ is a vector of the reciprocal lattice. 
\item if one of the components $\rec_1, \rec_2$ or $\rec_2-\rec_1$ is an \emph{even integer}, and in particular, if $\brec$ is collinear with one of the basis vectors $\bs b_{1,2}$. For example, when $\rec_1=0$ (resp. $\rec_2=0$), one finds $\Phi_3=0$ and $\Phi_1=-\Phi_2$ (resp. $\Phi_2=0$ and $\Phi_1=-\Phi_3$).
\end{itemize}
In these pathological ``staggered flux" cases, the system remains invariant under time reversal and therefore topologically trivial (note that the number of magnetic flux quanta $\Phi_{\alpha}$ is only defined modulo 1).  

We can verify that these singular time-reversal configurations equally correspond to the condition 
\be
d_z (\bs K_{+})=d_z (\bs K_{-}), \forall t_A,t_B. \label{condtr}
\ee 
As established in Eq.~(\ref{chern2}), the condition \eqref{condtr} naturally leads to a trivial Chern insulator $\nu=0$ when $\Delta > 0$, as expected for a time-reversal-invariant system exhibiting a gap. One can check that the condition \eqref{condtr} can be simply rewritten in terms of the vector $\brec = \rec_1 \bs b_1 + \rec_2 \bs b_2$,  
\be
\sin(\pi\rec_2)-\sin(\pi\rec_1)+\sin\left[ \pi(\rec_2-\rec_1)\right]=0, \label{condtr2}
\ee
whose solutions exactly reproduce the pathological cases listed above. When $\brec$ satisfies the condition \eqref{condtr2}, the system is necessarily a trivial insulator or a standard semi-metal depending on the other  parameters. In Section~\ref{deltap}, we explore other values of the momentum recoil $\brec$, where trivial or non-trivial phases can be found depending on the specific values of the parameters $t_A,t_B,\varepsilon$. \\

\subsection{Comparison with the Haldane model}
\label{haldanesection}

We conclude this Section by comparing the laser-coupled honeycomb lattice \eqref{htot}, with the original Haldane model (cf. \cite{Haldane1988} and Fig. \ref{figfluxlattice} (b)). In the latter, the hopping factor $t_1$ between NN sites of the honeycomb lattice is real, while NNN hoppings $t_2$ are multiplied by a constant phase factor $e^{ \pm i 2 \pi \phi_{\text{H}}}$ (the sign being determined by the orientation of the path). Thus, in the Haldane model, the three small triangular subplaquettes illustrated in Fig. \ref{figfluxlattice} (b) are all penetrated by the \emph{same} flux $\Phi_{1,2,3}=\phi_{\text{H}}$, whereas the large central triangular plaquette is penetrated by a flux $- 3 \phi_{\text{H}}$. This leads to a staggered magnetic field configuration, with a vanishing total flux penetrating the hexagonal unit cell $\Phi(\hexagon)=0$. We stress that time-reversal symmetry is \emph{necessarily broken} in the Haldane model, for \emph{any} finite value of the phase $\phi_{\text{H}} \ne 0$. This important difference between the two models highlights the richness of the laser-coupled honeycomb lattice \eqref{hamtwo}-\eqref{sz}, where the flux configuration and the nature of the spectral gaps strongly depend on  the orientation of the vector $\brec$ entering the Peierls phases.

\section{Phase diagrams for topological insulating phases}
\label{topologicalphases}

In this section, we perform a systematic characterization of the phase diagram. We set the nearest-neighbor tunneling amplitude to $t =1$, thus effectively
measuring all energies in units of $t$. The Cartesian components of the recoil momentum $p_x$ and $p_y$ are conveniently measured in units of $K_x$ and $K_y$, which are the coordinates of the
Dirac point ${\bs K}_{+}$, with $K_x = 2\pi/3$ and $K_y = 2\pi / 3\sqrt{3}$. Following the discussions in the preceding Section, we can expect three different phases :
\begin{itemize}
\item a semi-metal (energy gap $\Delta=0$),
\item an insulator (energy gap $\Delta \neq0$) with trivial topology ($\nu=0$),
\item a Chern insulator ($\Delta\neq0, \nu\neq0$).
\end{itemize}

At this point, let us remind ourselves that the Chern number $\nu$ defined in Eq. \eqref{chern} characterizes the topological order of \emph{insulating} phases \cite{Kohmoto:1985}. However, the expression in Eq. \eqref{chern2} could also be formally computed for a semi-metal configuration ($\Delta =0$), but in this case, the index $\nu$ cannot be associated with a robust and topologically protected Hall conductivity. This fact, which is crucial from the experimental detection point of view, is further elaborated in the next Section \ref{skyrmion}. In this Section, where the focus is set on Chern insulators, we are therefore looking for wide regions in parameter space where both $\Delta$ and $\nu$ are non-zero. In Section \ref{deltap}, we consider how the system evolves as the recoil momentum $\brec$ is varied without staggered potential ($\varepsilon=0$). We examine further the role of anisotropy in the tunneling energies ($t_A \neq t_B$) in Section \ref{anisotropysection}, and finally the role of a  staggered potential ($\varepsilon\neq 0$)  in Section \ref{staggeredsection}.

\subsection{Recoil momentum}\label{deltap}

\begin{figure}
\centering
\includegraphics[width=1.\columnwidth]{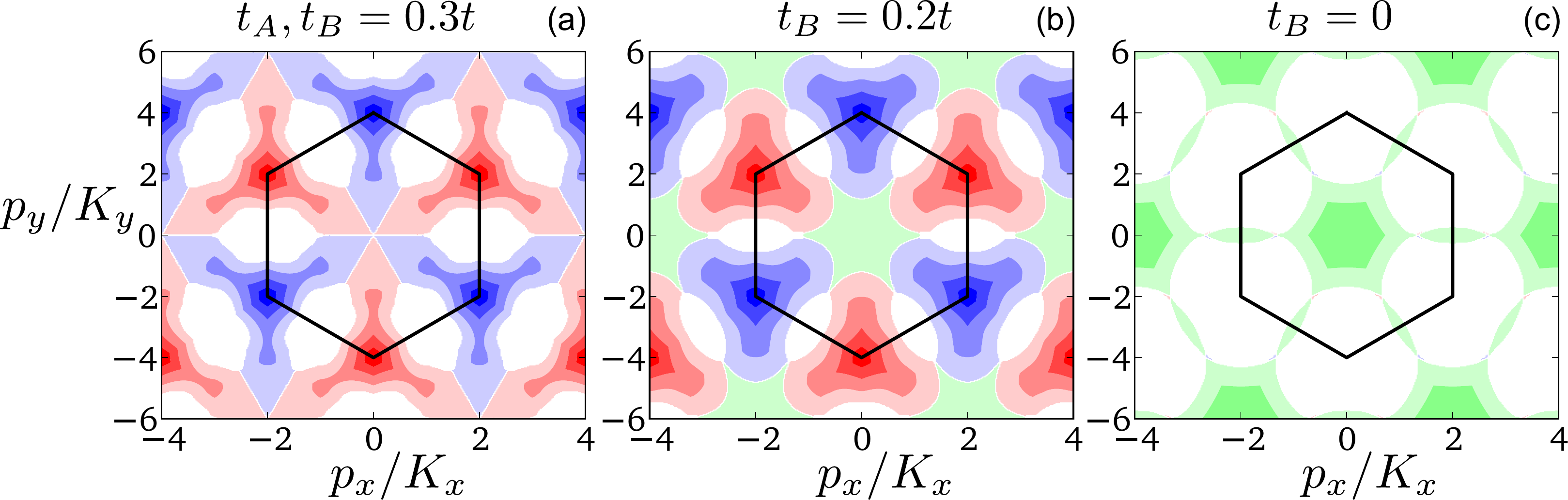}
\caption{\label{figa1} Phase diagrams as a function of the recoil momentum
components $p_x$ and $p_y$. In all the figures, we set $t=1$,
$\varepsilon=0$ and $t_A=0.3 t$. In panel (a) we set $t_B = t_A = 0.3 t$,
and in panel (b) $t_B = 0.2 t$. The extreme case $t_B=0$ is shown in (c).
The white regions correspond to metallic phases (i.e. vanishing of the gap $\Delta \approx 0$), the
blue and red regions correspond to topological phases with $\nu=\pm 1$.
The green regions correspond to trivial insulating phases $\nu=0$. The ``resized''
FBZ is indicated by a hexagon, which also serves to highlight the angle
dependence with respect to the inverse lattice vectors. The size of the gaps is indicated by the intensity: the lightest shades denote areas where the gaps are $\Delta <0.1 t$ and the brightest areas correspond to $1.5t < \Delta < 2 t$.} 
\end{figure}

We first investigate the effects of the Raman recoil momentum $\brec$. Here, the staggered potential is set to
$\varepsilon = 0$. The phase diagrams shown in Fig.~\ref{figa1} illustrate
the appearance of topological phases as a function of the Cartesian components
$p_x$ and $p_y$, for several values of the tunneling rates $t_{A,B}$. The areas corresponding to nontrivial topological phases, characterized by the Chern numbers $\nu = \pm 1$, are indicated by blue and red colors,
respectively. Green areas correspond to the trivial insulating phase $\nu=0$, and
white areas signify the ``undesired" metallic regime ($\Delta \approx 0$). The size of the bulk gap $\Delta$ is simultaneously shown through the color intensity. Panel (a) shows the isotropic case with equal next-nearest neighbor tunneling amplitudes set to $t_B = t_A = 0.3 t$. Here, non-trivial
topological phases ($\nu= \pm 1$) are generally separated by semi-metallic
or metallic phases, and these topological regions depict triangular patterns. Panels (b) and (c) correspond to anisotropic cases where
the hopping amplitude $t_B$ is reduced to $t_B = 0.2 t$ in panel (b) and set to zero in panel (c). As the anisotropy increases, the metallic regions and trivial insulating phases progressively modify the non-trivial islands.

In the special case $t_B=0$, we find that all the regions that were non-trivial for $t_B > 0$ reduce to semi-metals: when $t_B=0$, no topological insulating phase is found (contrary to what the Skyrmion behavior of the vector $\bs d$ would suggest \cite{Alba2011}, see Section \ref{skyrmion}). We stress that the semi-metallic behavior of the special case $t_B=0$ is found for the entire parameter space (i.e. for all $t_A$, $\varepsilon$ and $\brec$), and equally happens for the case $t_A=0$ and $t_B\ne0$. This subtle effect is highlighted in Fig. \ref{figa7} (c), presented in Section \ref{anisotropysection}, where the band structure $E=E(k_y)$ clearly shows the indirect gap closing for the case $t_B=0$. This energy spectrum suggests that a small perturbation could open the bulk gap and lead to a Chern insulator. However, in Section \ref{staggeredsection}, we show that the staggered potential does not open such a non-trivial gap in the case $t_B=0$. We therefore conclude that the condition $t_{A,B} \ne 0$ should be satisfied to generate a robust Chern insulator. 

The Hamiltonian is a periodic function of $\brec$, and the resulting
periodicity of phases is conspicuous in the phase diagrams illustrated in Fig.~\ref{figa1}. The central
elementary lattice (the ``resized" FBZ) cell is marked by a black hexagon \footnote{To be more precise, the arguments of the cosines and the complex exponentials in the Hamiltonian feature $\brec / 2$, thus
the ``resized" FBZ is twice larger than the actual FBZ. The panels of Fig.~\ref{figa1} show
rectangular regions containing exactly four Brillouin zones.} in all
panels of Fig.~\ref{figa1}. We find that the most convenient non-trivial topological insulating phases (i.e. phases protected by the largest bulk gaps $\Delta \sim 2 t$) are found for $\brec \propto (\sin N \pi/3 , \cos N \pi/3 )$, where $N$ is an integer. Therefore, setting $p_x=0$ potentially leads to topological phases with large bulk gaps, which is the most interesting situation for an experimental realization (cf. Section \ref{skyrmion}).

\begin{figure}
\centering
\includegraphics[width=0.5\columnwidth]{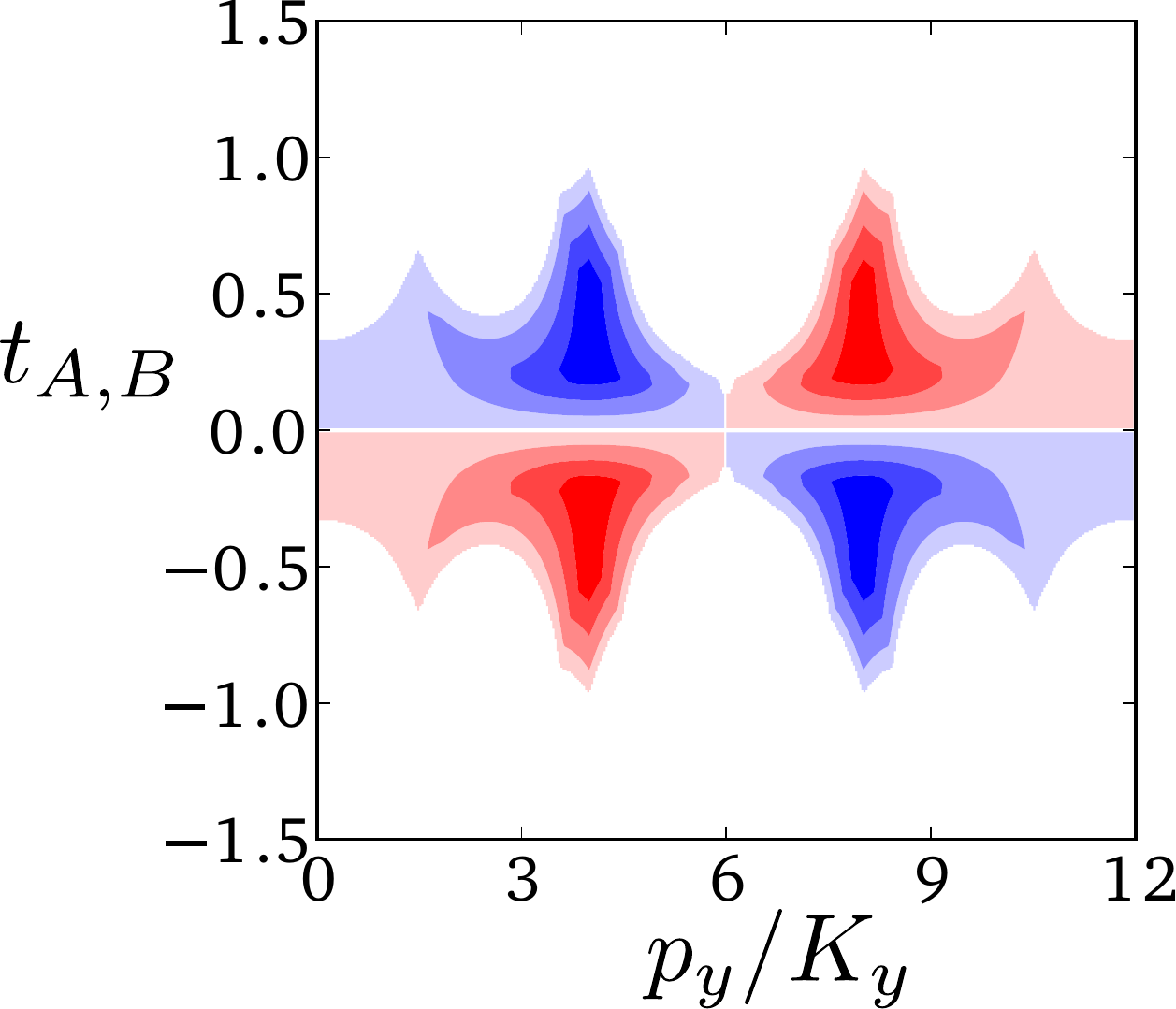}
\caption{\label{figa2} Phase diagrams as a function of the tunneling amplitude $t_A$ and the recoil momentum component $p_y$. We set $t=1$, $\varepsilon=0$, $\brec=(0,p_y)$ and $t_A = t_B$. The color code is the same as in Fig. \ref{figa1}.} 
\end{figure}

We now explore how the topological phases evolve as the laser recoil
momentum $\bs{p}$ and the tunneling amplitudes $t_{A,B}$ are modified. As motivated above, 
we set $p_x=0$, and then compute the phase diagrams in the
$p_y - t_A$ plane. First of all, we investigate the isotropic case $t_A=t_B$ (the effects of anisotropy will be discussed in Section \ref{anisotropysection}). The phase diagram presented in Fig.~\ref{figa2}
indicates that in the realistic situation where $t_A \approx t_B$, the sizes of the topological gaps $\Delta$ are maximum for $t_A \approx t_B \approx 0.3 t$, where $\Delta \approx 2 t$ for $p_y \approx 4 K_y$ (see also Fig. \ref{figa1} (a)). Furthermore, this figure indicates that one should generally observe phase transitions between 
metallic and non-trivial topological phases as $p_y$ is varied. Importantly, we note that the system remains metallic ($\Delta=0$) when the ``natural" hoppings $t_{A,B}$ are \emph{larger} than the Raman-induced hopping $t$, in particular when $t_A \approx t_B \approx t$. In the following, we show that an anisotropy $t_A \ne t_B$ , or the inclusion of a staggered potential $\varepsilon \ne 0$, can turn this metallic phase into a topological one.

\subsection{Anisotropy}
\label{anisotropysection}

\begin{figure}
\centering
\includegraphics[width=0.5\columnwidth]{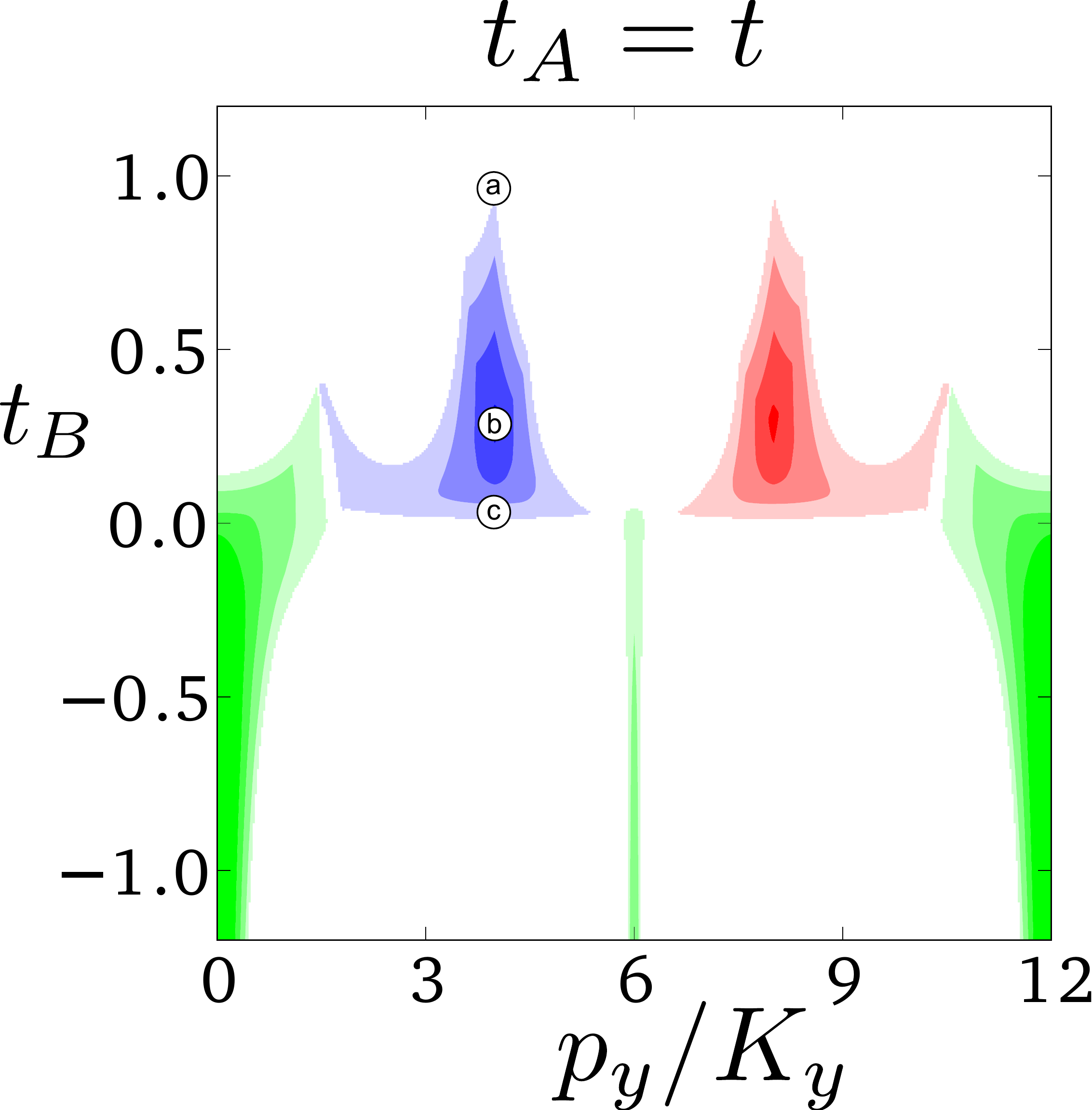}
\caption{\label{figa3} Phase diagram as a function of the tunneling
amplitude $t_B$ and the recoil momentum component $p_y$, as $t_A=t$ is
fixed.  Here, we set $t=1$, $\varepsilon=0$ and $p_x=0$. The color code is the same as in Fig. \ref{figa1}. The three dotted configurations (a)-(c) are further illustrated through band structures $E=E(k_y)$ in Figs. \ref{figa7} (a)-(c). } 
\end{figure}

\begin{figure}
\centering
\includegraphics[width=1.02\columnwidth]{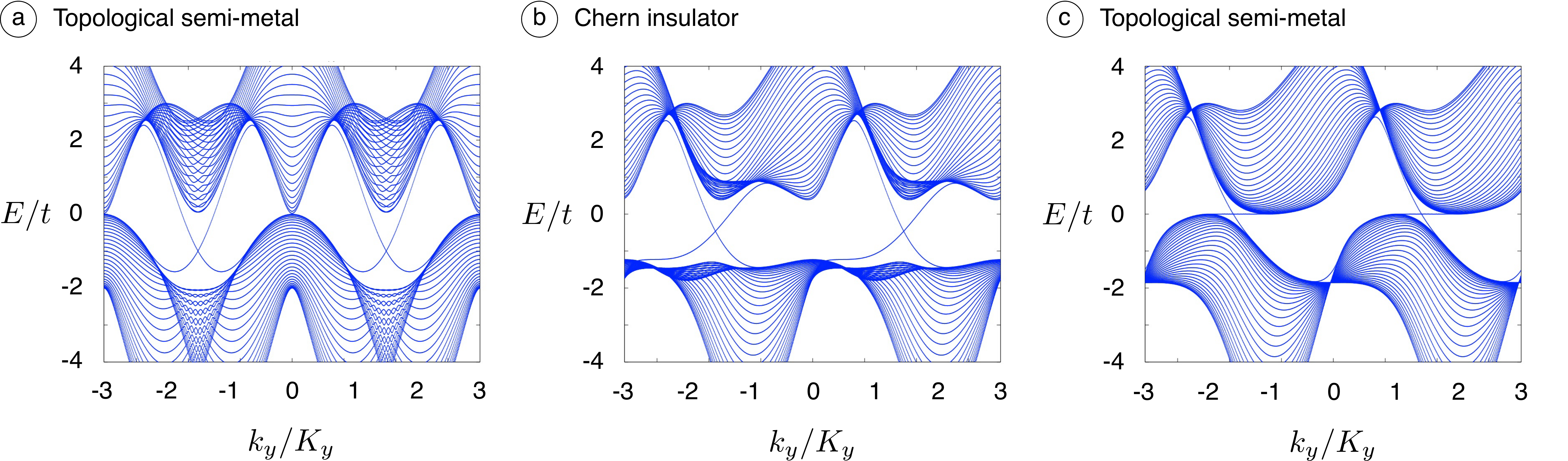}
\caption{\label{figa7} Energy spectra $E=E(k_y)$, as a function of the quasi-momentum $k_y$, for a cylindrical geometry with zigzag edges. The parameters in (a)-(c) correspond to the configurations labelled by dots in Fig. \ref{figa3}: namely, $\varepsilon=0$ and (a) $t_A=t_B=t$, (b) $t_B=0.3 t$, (c)  $t_B=0$. In all the figures $t_A=t=1$, $\varepsilon=0$ and $p_y= 4 K_y$. When $\nu\ne 0$ and $\Delta \ne 0$, as in Fig. (b), gapless dispersion branches cross the bulk energy gap: they describe current-carrying edge states, which lead to a quantized Hall conductivity \cite{Hatsugai1993}. Figs. (a),(c) illustrate the peculiar situations where the gap indirectly closes, $\Delta = 0$, and where the winding number \eqref{qiformula} is non-trivial $w \ne 0$ (cf. Section \ref{semimetal}).} 
\end{figure}

In Fig.~\ref{figa3}, we show the phase diagram in the plane $p_y - t_B$
for a large and fixed value of the tunneling rate $t_A=t$. This important
result shows that when $t_A \approx t$, the anisotropy $\vert t_A - t_B \vert \ne 0$
is necessary to open non-trivial topological gaps. This effect occurs for
a relatively large range of the anisotropy, namely for
$t_B \in  ] 0 ,  t_A]$, and for specific values of the momentum $p_y$.
For larger anisotropy $\vert t_A - t_B \vert > t$, the topological
phases are destroyed and only metallic and trivial insulators survive. Specific phase transitions between semi-metallic and Chern insulating phases, indicated in Fig.~\ref{figa3} by three successive dots, are further illustrated through the edge-state analysis, in Fig.~\ref{figa7}. In panel \ref{figa7} (b), one indeed observes the presence of topological edge states within the bulk gap, which is the hallmark of a Chern insulator, i.e. through the bulk-edge correspondence \cite{Hatsugai1993}. Finally, we note the robustness of the topological edge states within the semi-metal regime  $\Delta =0$, in Figs. \ref{figa7} (a),(c), a fact which is further analyzed in Section \ref{skyrmion}.

\subsection{Staggered potential}
\label{staggeredsection}

\begin{figure}
\centering
\includegraphics[width=\columnwidth]{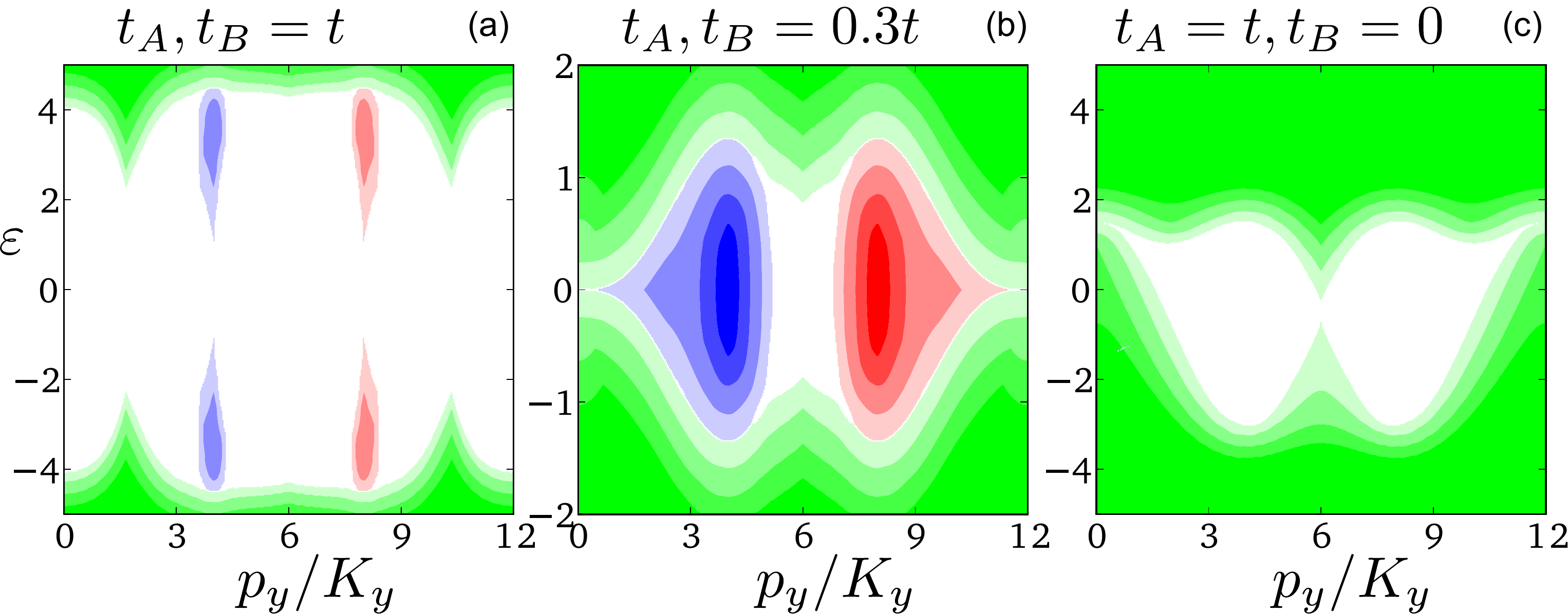}
\caption{\label{figa4} Phase diagrams as a function of the staggered 
potential strength $\varepsilon$ and the recoil momentum component $p_y$. In all the figures, we set $t=1$ and $p_x=0$. (a)-(b) The isotropic cases $t_A=t_B=t$ and $t_A=t_B=0.3t$. The extreme case where $t_B=0$ is shown in (c). The color code is the same as in Fig. \ref{figa1}.} 
\end{figure}

In this Section, we explore the effect of the staggered potential. In 
Fig.~\ref{figa4}, we show the phase diagram as a function of the staggered 
potential strength $\varepsilon$ and of the recoil momentum component $p_y$, for 
several configurations of the tunneling amplitudes $t_{A,B}$ (we set 
$p_x=0$). First, we show the case $t_A=t_B=t$ in Fig.~\ref{figa4} (a). 
In this situation, large metallic regions and small non-trivial islands are found in the phase diagram, which can already be anticipated from Fig.~\ref{figa2} for $\varepsilon=0$. Interestingly, in the totally symmetric case, where $t_A=t_B=t$, the topological phases vanish for $\varepsilon=0$, and they are thus separated along the $\varepsilon$ axis\footnote{Note that the vanishing of 
the topological insulating phases for $t_A=t_B=t$ and $\varepsilon=0$ can be 
visualized in Fig.~\ref{figa2}.}.
These results indicate that the 
staggered potential is \emph{necessary} to induce topological phases in this 
situation where $t_A=t_B=t$. However, for large values of the staggered potential, a trivial phase 
with $\nu=0$ is always privileged, in agreement with the general belief that 
such a staggered potential generically leads to trivial phases \cite{Haldane1988}.   

For $t_A=t_B<t$, non-trivial Chern insulating phases can be formed both with and without the staggered potential. In Fig.~\ref{figa4} (b), we illustrate the effects of the staggered potential for the optimized values of the tunneling rates $t_A= t_B=0.3 t$. Here, one observes two topological phases with $\nu=\pm 1$, which are separated by a small metallic region. This result highlights that one should generally observe phase transitions between semi-metallic and non-trivial topological phases as $p_y$ is varied. On the other hand, varying the staggered potential to large values always privileges the transition to a trivial phase with $\nu=0$. 

In the extreme case where $t_B=0$, shown in Fig.~\ref{figa4} (c), one finds the \emph{contour} of the phase diagram presented in Ref. \cite{Alba2011}. However, we stress that the two central regions featured in this diagram do not correspond to Chern insulating phases, as their corresponding bulk gap is \emph{closed}. This indirect gap closing is further illustrated in Fig. \ref{figa7} (c). In the next Section, we analyze this important point in more details.

\section{The winding number and the ToF measurement: Chern insulators, topological semi-metals and Skyrmions}\label{skyrmion}

In the previous Section, we have identified the topological insulating phases that could be realized in our cold-atom system, when the parameters $(t_{A,B}, \varepsilon, \bs p)$ are tuned in the gapped regimes $\Delta > 0$. In these situations, the Chern number \eqref{chern} associated with the low-energy eigenstate $\vert u_{-} \rangle$ can be defined, and its experimental measure would witness a clear manifestation of non-trivial topological order. However, contrary to solid-state experiments where the Chern number is directly evaluated through a Hall conductivity measurement \cite{vonKlitzing:1986}, it can only be observed indirectly in the cold-atom framework \cite{Alba2011,Hafezi:2007,Palmer:2008,umucalilar2008a,Bermudez:2010,Stanescu:2010,Price2012,Goldman2012}. In this Section, we analyze in detail the topological orders which could be detected through a ToF experiment \cite{Alba2011}, and further discuss the role played by the bulk gap $\Delta$ in this context. \\  

First of all, let us note that the Hamiltonian \eqref{hamtwo} can be associated with a topological (Pontryagin) winding number
\cite{Qi:2006,Girvin:1999,Konig:2008,Prada:2011},
\begin{align}
 w &= \frac{1}{4 \pi} \int_{\mathbb{T}^2} \bs n \cdot \biggl ( \partial_{k_x} \bs n \times  \partial_{k_y} \bs n  \biggr) \txt{d}^2 \bs{k}, \nonumber \\
 &= \frac{1}{4 \pi} \int_{\mathbb{T}^2} \frac{\bs{d}}{d^3} \,   \cdot \biggl ( \partial_{k_x} \bs{d} \times  \partial_{k_y} \bs{d}    \biggr) \txt{d}^2 \bs{k}, \label{qiformula}
\end{align}
which measures the number of times the unit vector $\bs n (\bs k)=\bs{d} (\bs k)/d (\bs k)$ covers the Bloch sphere $S^2$ as $\bs k$ evolves on the entire FBZ \cite{Qi:2006}.   When $w \ne 0$, this leads to a Skyrmion configuration for the vector field $\bs{n}(\bs k)$.   As will be discussed later in this Section and depicted in Fig. \ref{skyrmionfig}, the Skyrmion configuration corresponds to a situation where the unit vector $\bs{n} (\bs{k})$ entirely covers the Bloch sphere once, which for the present model implies that the vector $\bs{n} (\bs{k})$ points in opposite directions (i.e. North and South poles) at the two inequivalent Dirac points, 
\begin{eqnarray}
&w=+1 \longrightarrow \bs n (\bs K_{+})= + \bs{1}_z \text{ and } \bs n (\bs K_{-})= - \bs{1}_z, \nonumber \\
&w=-1 \longrightarrow \bs n (\bs K_{+})= - \bs{1}_z \text{ and } \bs n (\bs K_{-})= + \bs{1}_z. \label{poleswinding}
\end{eqnarray}
The winding number $w$ characterizes the map $\bs n (\bs k) : \mathbb{T}^2 \rightarrow S^2$ defined in Eq. \eqref{sz}, and therefore, it is not necessarily related to the spectrum or eigenstates of the Hamiltonian  \eqref{hamtwo} -- contrary to the Chern number  \eqref{chern}, which is a mathematical index associated with the state $\vert u_{-} \rangle$ \cite{nakahara}. \\

In this work, a \emph{topological semi metal} denotes a gapless phase $\Delta =0$, characterized by a non-trivial winding number $w\ne 0$. The fate of the winding number $w$ and its corresponding Skyrmion pattern will be discussed in Subsection \ref{semimetal}, where these structures are shown to remain stable when $\Delta =0$, as long as the gap does not close at the Dirac points. In fact, when the gap is open $\Delta > 0$, the Chern number \eqref{chern} is exactly equal to the winding number \eqref{qiformula}, \footnote{The Chern number \eqref{chern} and its corresponding fibre bundle structure \cite{nakahara} could also be formally defined when the gap is indirectly closed, such as in Fig. 6 (c). Thus, the Chern number $\nu$ and the winding number $w$ are formally equivalent under the more general gap-opening condition \cite{Hatsugai:2005}: $E_-(\bs k) < E_{+} (\bs k)$ for all $\bs k \in \text{FBZ}$. In the present model, this condition reads $d_z(\bs K_{\pm}) \ne 0$.}
\be
 \nu = w , \label{chernwind2}
\ee 
as can be demonstrated using Eqs. \eqref{chernone},\eqref{A-pm} and \eqref{pm}  (cf. also Refs. \cite{Qi:2006,Konig:2008,Prada:2011}). As a corollary, the result in Eq. \eqref{poleswinding} can be easily deduced from Eq. \eqref{chern2}. From the equivalence \eqref{chernwind2}, we observe that the Chern insulating phases discussed in the previous Sections are characterized by a non-trivial winding number $w \ne 0$, and therefore, they are also associated with a Skyrmion pattern. In summary, measuring the winding number $w$ in an experiment would allow to equally identify Chern insulators ($\Delta >0$) and topological semi-metals ($\Delta = 0$).\\

As first observed in Ref. \cite{Alba2011}, the vector field $\bs n (\bs k)$ could be detected through a ToF absorption image. From such data, one could then evaluate the winding number $w$, using a discretized version of Eq. \eqref{qiformula}. This detection method is based on the fact that $\bs n (\bs k)$ can be expressed in terms of the momentum densities $\rho_{A,B} (\bs k)$ associated with the two spin species $A,B$ (cf. Fig. \ref{figlattice}). Defining the regions $\mathcal{K}_{(\pm)}=\{ \bs k : E^{(\pm)}(\bs k) < E_F \}$, we find that 
\begin{align}
\rho_B (\bs k) - \rho_A (\bs k) &= + n_z (\bs k) & &\text{for $\bs k \in \mathcal{K}_{(-)}$ and  $\bs k \notin \mathcal{K}_{(+)}$}, \nonumber \\
\rho_B (\bs k) - \rho_A (\bs k) &= - n_z (\bs k) & &\text{for $\bs k \in \mathcal{K}_{(+)}$  and $\bs k \notin \mathcal{K}_{(-)}$}, \nonumber \\
\rho_B (\bs k) - \rho_A (\bs k) &=0 & &\text{otherwise} \nonumber.
\end{align}
Unfortunately, one cannot generally determine the regions $\mathcal{K}_{(\pm)}$ in an experiment, unless the Fermi energy is exactly located in a bulk gap (in which case $\mathcal{K}_{(-)}=\text{FBZ}$ and $\mathcal{K}_{(+)}=\emptyset$). Therefore, the vector field $\bs n (\bs k)$ can only be approximately reconstructed from the data when both bands $E_{\pm} (\bs k)$ are partially filled. In fact, if we apply the relation $\rho_B (\bs k) - \rho_A (\bs k) = n_z (\bs k)$ to every pixel of a ToF image \footnote{The other components $n_{x,y} (\bs k)$ of the vector field could be obtained through similar measurements, combined with a rotation of the atomic states \cite{Alba2011}. }, and discretize the expression \eqref{qiformula} to evaluate the winding number from this data, we would experimentally measure the following quantity
\begin{align}
 w_{\text{ToF}}&= \frac{1}{4 \pi} \Biggl ( \sum_{ \mathcal{K}_{(-)}} - \sum_{ \mathcal{K}_{(+)}} \Biggr )
 \sum_{\mu \ne \nu \ne \lambda}  n_{\mu} (\bs k) \biggl (  n_{\nu} (\bs k + \bs e_x) n_{\lambda} (\bs k + \bs e_y) - n_{\nu} (\bs k +  \bs e_y) n_{\lambda} (\bs k +  \bs e_x) \nonumber \\
 &+ n_{\nu} (\bs k + \bs e_y) n_{\lambda} (\bs k)  - n_{\nu} (\bs k +  \bs e_x) n_{\lambda} (\bs k) + n_{\nu} (\bs k) n_{\lambda} (\bs k +  \bs e_x)  - n_{\nu} (\bs k) n_{\lambda} (\bs k +  \bs e_y)    \biggr ),\label{dis2}
 \end{align}
where $\mu,\nu,\lambda=x,y,z$, and where $\bs e_{x,y}$ are the two unit vectors defined on the discretized FBZ. When the Fermi energy is set within a bulk gap, only the first sum contributes $\sum_{ \mathcal{K}_{(-)}}=\sum_{\text{FBZ}} $, and the quantity $w_{\text{ToF}}$ converges towards the winding number $w$ as the resolution of the grid is increased (cf. \ref{app:finite}). When the gap is closed, and if the Fermi energy is tuned such that the bulk bands $E_{\pm} (\bs k)$ are only partially filled, the quantity $w_{\text{ToF}}$ will generally deviate from the quantized value $w$. Consequently, the assumption of a perfectly filled lowest band ($\mathcal{K}_{(-)}=\text{FBZ}$ and $\mathcal{K}_{(+)}=\emptyset$), as considered in the calculations of Ref. \cite{Alba2011}, is crucial in the case $\Delta =0$. However, we indicate that this condition would be difficult to fulfill in an experiment, due to experimental imperfections and finite temperatures. We now illustrate this discussion in Subsections \ref{chernwind}-\ref{semimetal}, where the signatures of Chern insulators and topological semi-metals are compared and commented. Let us finally remark that the quantity $w_{\text{ToF}}$ defined in Eq. \eqref{dis2} is strictly equivalent to the discretized expression for the Hall conductivity $\sigma_H$, which is not necessarily quantized in the general case where the Fermi energy is not located in a bulk gap (cf. \ref{app:hall}).

\subsection{The Chern insulators}
\label{chernwind}
\begin{figure}
\centering
\includegraphics[width=0.7\columnwidth]{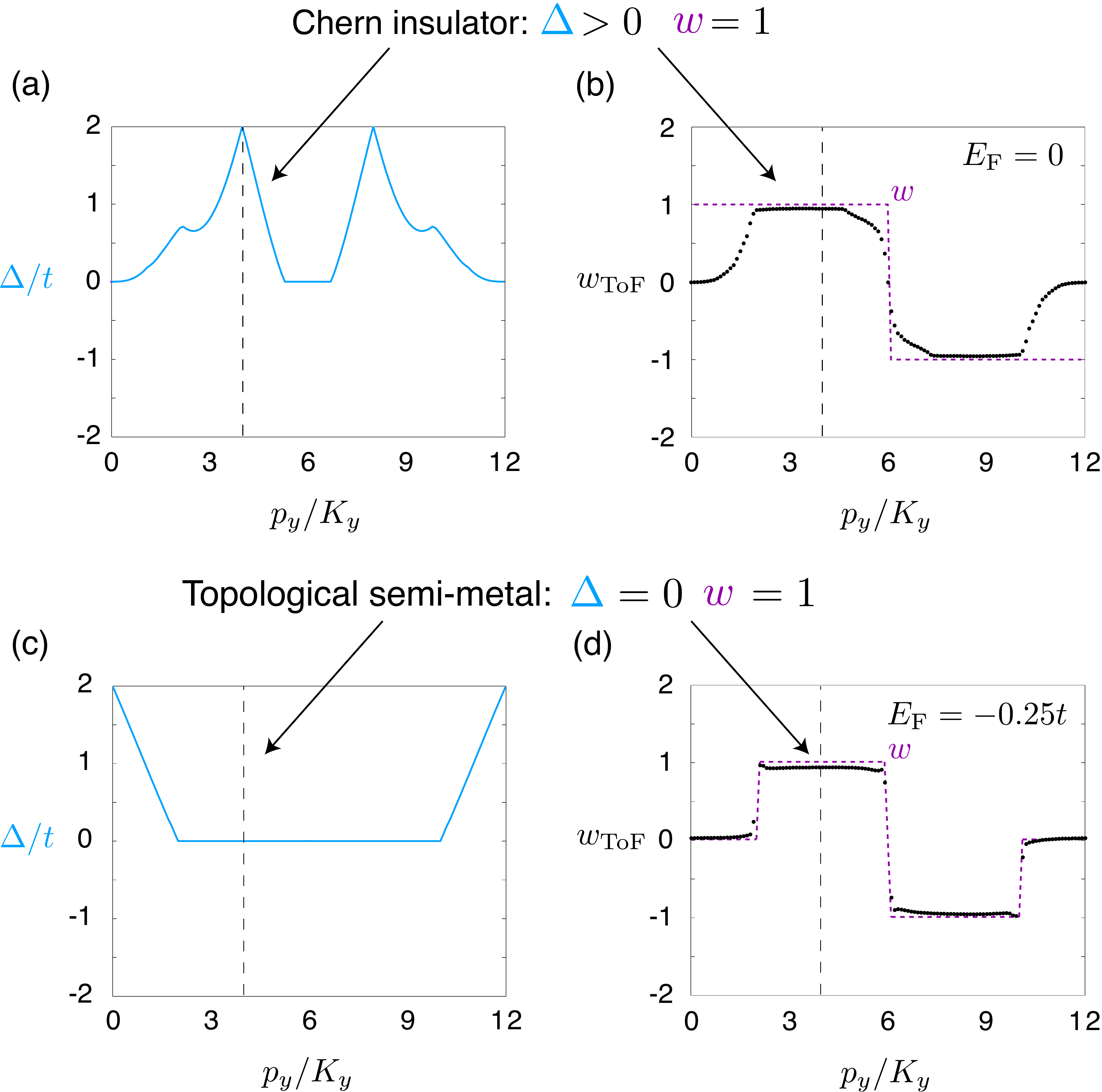}
\caption{The energy gap $\Delta$ and the discretized winding number $w_{\text{ToF}}$ as a function of $p_y$ for (a)-(b) $t_A=t_B=0.3 t$, $\varepsilon=0$ and (c-d)  $t_A=0.5 t$, $t_B=0$, $\varepsilon=-0.5 t_A$. For all plots $p_x=0$. The discretized winding number $w_{\text{ToF}}$ has been computed from Eq.\eqref{dis2} using a $30 \times 90$ lattice and setting the Fermi energy: (b) $E_{\text{F}}=0$ (i.e. inside the gap); (d) $E_{\text{F}}=-0.25$ (i.e. at the gap closing point). For comparison, purple dotted lines show the integral winding number $w$ defined in Eq. \eqref{qiformula}. The parameters in (c)-(d) are the same as in Ref. \cite{Alba2011}. In all the figures, a vertical dashed line shows the value $p_y=4 K_y$ used in Fig. \ref{gaptopology2}.} \label{gaptopology}
\end{figure}

\begin{figure}
\centering
\includegraphics[width=0.7\columnwidth]{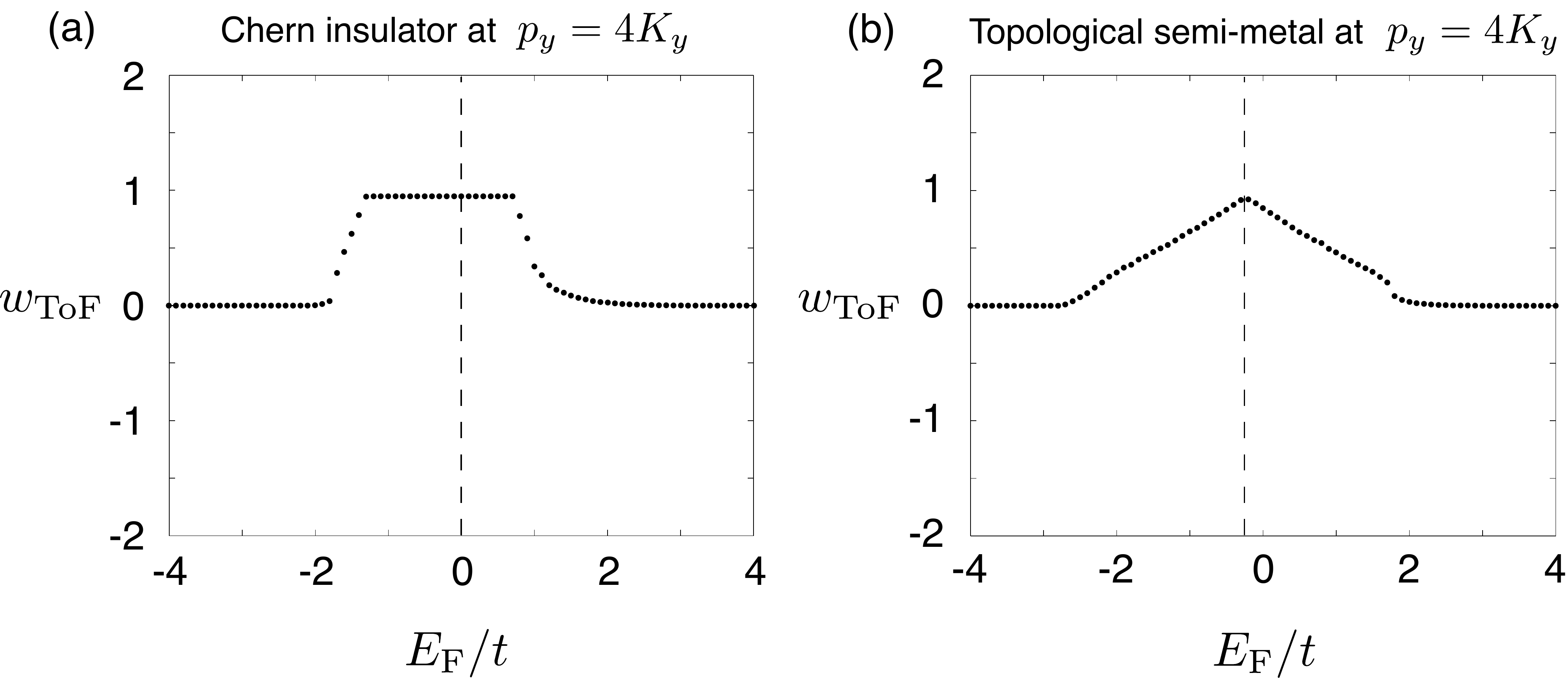}
\caption{The winding number $w_{\text{ToF}}$  as a function of the Fermi energy $E_{\text{F}}$ at zero temperature for $p_x=0,p_y=4 K_y$. The parameters in (a) are the same as in Figs.~\ref{gaptopology}a-b, whereas the parameters in (b) are the same as in Figs.~\ref{gaptopology} c-d. The computations were performed using Eq.\eqref{dis2}, on a $30 \times 90$ lattice. The dashed lines show the values (a) $E_{\text{F}}=0$, (b) $E_{\text{F}}=-0.25$ used in Figs. \ref{gaptopology} (b) and (d), respectively.} \label{gaptopology2}
\end{figure}

When a spectral gap is opened, $\Delta >0$, the winding number \eqref{qiformula} is exactly equal to the Chern number \eqref{chernone}. This potentially gives rise to a Chern insulator, as illustrated in Figs. \ref{gaptopology} (a)-(b), where we compare how the energy gap $\Delta$, the winding number $w$ and the ToF measurement $w_{\text{ToF}}$ vary as a function of the recoil momentum $p_y$. As expected from the topological property of the Chern number, we find that the ranges where $\Delta >0$ and $\nu =w= \pm 1$ lead to the clear plateaus depicted by the observable $w_{\text{ToF}}(p_y) \approx \pm 1$ (cf. \ref{app:finite} for a discussion on finite size effects). We also demonstrate the robustness of these plateaus in Fig. \ref{gaptopology2} (a), where the quantity $w_{\text{ToF}}$ is computed as a function of the Fermi energy $E_{\text{F}}$, and where a large plateau $\sim \Delta$ is observed for fixed values of the other parameters. Therefore, the ToF winding number $w_{\text{ToF}}$ shows a robust behavior, and exhibits a clear plateau when the Fermi energy lies in the band gap. In other words, the Chern insulating phase is characterized by a ``quantized" winding number $w_{\text{ToF}}$, which is protected against finite changes of the parameters through the existence of a topological bulk gap \footnote{The plateaus depicted by $w_{\text{ToF}}$ are strictly equivalent to Hall conductivity plateaus (cf. \ref{app:hall}). However, since the Hall conductivity is not measured in cold-atom experiments, we choose to represent the observable quantity $w_{\text{ToF}}$ in our plots, rather than $\sigma_H$.}. \\

Let us stress the important fact that the Chern number $\nu$ in Eq. \eqref{chernone} no longer reflects the quantized Hall conductivity when the bulk gap is closed, in which case the system effectively describes a metallic phase. However, the topological order and Skyrmion patterns associated with the winding number $w$ survive even when the bulk gap is closed, as we now explore in the next Subsection.

\subsection{The topological semi-metals}
\label{semimetal}

\begin{figure}[!]
\centering
\includegraphics[width=1\columnwidth]{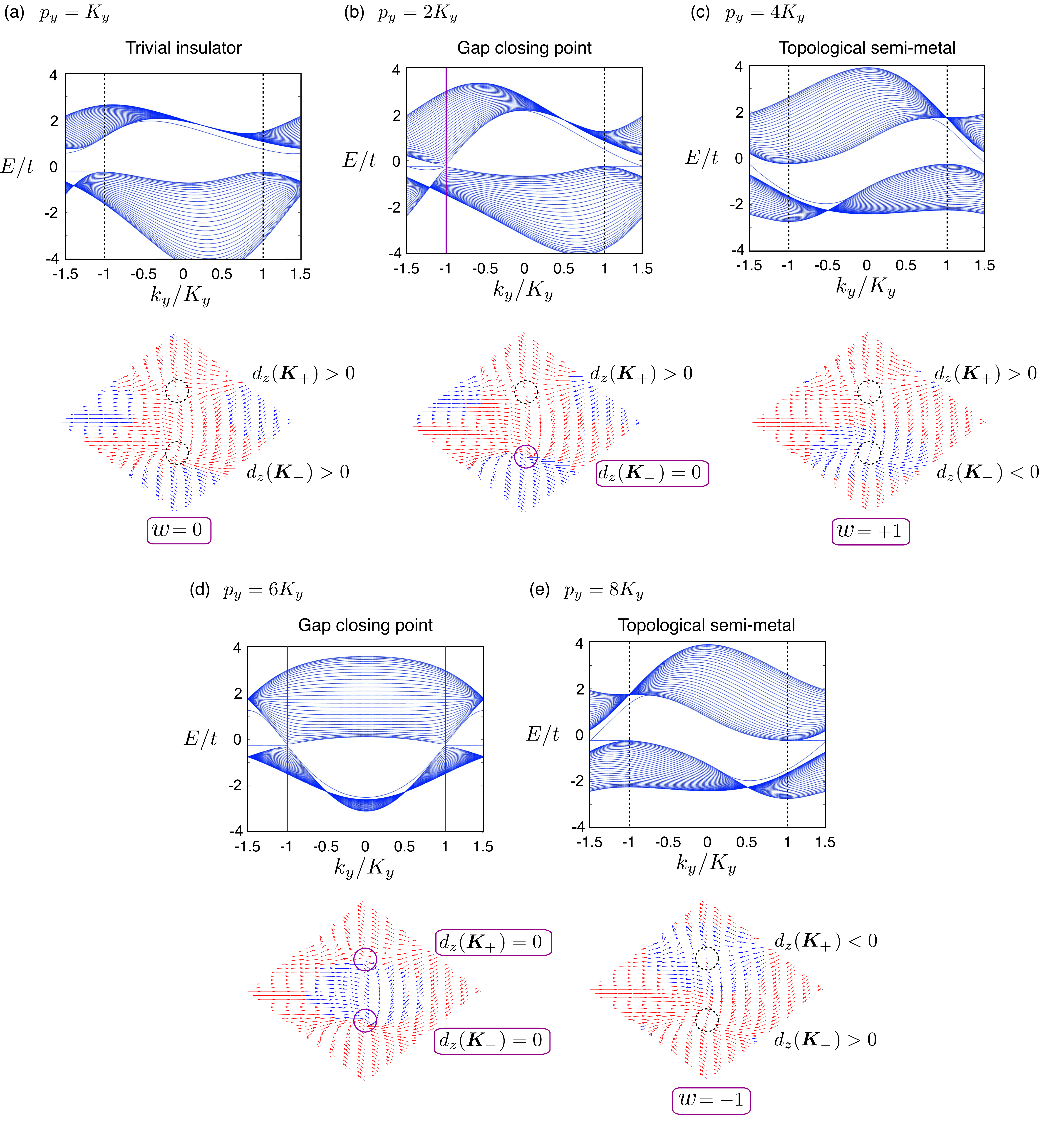}
\caption{Topological phase transitions: (top) Energy spectra $E(k_y)$ and (bottom) Skyrmion configuration depicted by the vector $\bs n (\bs k)$ within the FBZ. The parameters are the same as in Fig. \ref{gaptopology} (c)-(d), namely $t_A=0.5 t$, $t_B=0$ and $\varepsilon= - 0.5 t_A$. Note that the bulk gap is closed $\Delta = 0$ in all the figures except in (a). The $x$ and $y$ components of the normalized vector field $\bs{n}(\bs k)=(d_x (\bs k) , d_y (\bs k) , d_z (\bs k))/ d(\bs k)$ are represented by red arrows for $d_z (\bs k) > 0$ and blue arrows for $d_z (\bs k) < 0$. The corresponding winding numbers $w=0, \pm 1$ are also indicated. The location of the two Dirac points $\bs K_{\pm}$ are indicated by two vectical lines (top) and circles (below). A non-trivial winding number $w = \pm 1$ is clearly seen when the vector $\bs{n}(\bs k)$ has covered the whole Bloch sphere once.  Namely, when the North ($\bs n=+ \bs{1_z}$) and South ($\bs n = -\bs{1_z}$) poles have been reached at the two inequivalent Dirac points $\bs K_{\pm}$ (cf. Eqs.\eqref{chern2}-\eqref{chernwind2}). Note that topological phase transitions $w \rightarrow w'$ occur through a gap closing point, which is accompanied with the vanishing of $d_z (\bs K_D)$, at the Dirac point $\bs K_D$.} \label{skyrmionfig}
\end{figure}

First of all, we find that the ToF winding number $w_{\text{ToF}}$, given by Eq. \eqref{dis2}, can be robust \emph{even when the gap is closed}. This effect is illustrated in Figs. \ref{gaptopology} (c)-(d), where we compare how the energy gap $\Delta$, the winding number $w$ and the ToF winding number $w_{\text{ToF}}$ vary as a function of the recoil momentum $p_y$. From Figs. \ref{gaptopology} (c)-(d), we find that the winding number $w$ displays non-trivial plateaus $w=\pm 1$, in regions where the bulk energy gap is closed $\Delta = 0$. In Fig. \ref{gaptopology} (d), we precisely set the Fermi energy at the gap closing point $E_{\text{F}}=-0.25 t$, which can be determined from the spectra in Figs. \ref{skyrmionfig}(b)-(e). In this specific configuration, the observable winding number $w_{\text{ToF}}(p_y)$ depicts plateaus, and it converges towards the quantized value $w_{\text{ToF}} \rightarrow w$ as the resolution of the grid is increased (cf. \ref{app:finite}). However, as the Fermi energy is tuned away from this ideal value,  we find that the plateaus $w_{\text{ToF}}(p_y) \sim \pm 1$ progressively loose their robustness. This dramatic effect  is illustrated in Fig. \ref{gaptopology2}(b), where $w_{\text{ToF}}$ is computed as a function of the Fermi energy. Here, in contrast with the Chern insulator case shown in Fig. \ref{gaptopology2}(a), the winding number $w_{\text{ToF}}$ is strongly parameter-dependent: it only reaches $w_{\text{ToF}} \approx w=+1$ at the specific Fermi energy $E_{\text{F}}=-0.25 t$ (cf. Fig. \ref{gaptopology2} (b)). Consequently, the robust behavior of these topological semi-metals will only be observed if the Fermi energy is precisely tuned at the gap closing point (such as in Fig. \ref{gaptopology} (d)). This important fact makes topological semi metals more challenging to detect than Chern insulating phases. We point out that the phase diagrams and Skyrmion configurations presented in Ref. \cite{Alba2011}, and reproduced in Figs. \ref{gaptopology}(c)-(d), only feature trivial insulators and ``topological semi-metals", since all the computations were performed for the peculiar configuration $t_B=0$, whose corresponding spectrum remains gapless in the ``non-trivial" regions (cf. also Fig. \ref{skyrmionfig}). \\

The topological semi-metal is an intriguing phase, and in this new context, an important question arises: If the topological winding number $w$ remains stable for $\Delta = 0$ (cf. Fig. \ref{gaptopology} (d)), under which condition is this quantity changing its value? We address this question by analyzing the energy spectrum $E(k_y)$ together with the skyrmion pattern depicted by the vector $\bs n (\bs k)$ in Fig. \ref{skyrmionfig}. Here, the parameters are the same as in Fig. \ref{gaptopology} (d) and $p_y$ is varied between $K_y$ and $8 K_y$, where transitions between $w=0 \leftrightarrow w=+1$, but also between $w=+1 \leftrightarrow w=-1$, are expected. From Fig. \ref{skyrmionfig}, we find that the winding number $w$, and its corresponding Skyrmion pattern, remain extremely stable as long as the energy bands do not touch at the Dirac points. When a direct gap closing occurs at the Dirac point $\bs K_D$, where $\bs K_D$ denotes $\bs K_{+}$ and/or $\bs K_-$, we observe a topological phase transition signaled by a change in the winding number $w$ (cf. Fig. \ref{skyrmionfig} (b) and (d)). In particular, we find that this direct gap closing is  accompanied with the cancellation $d_z (\bs K_D)=0$ at the band touching point $\bs K_D$. In fact, the gap-closing condition $d_z (\bs K_D)=0$ should necessarily be satisfied at the transition between different values of the winding number $w$, as can be deduced from the equivalence \eqref{chernwind2} and from the simple expression \eqref{chern2}.  

The topological phase transitions are clearly visible on the Skyrmion patterns of Fig. \ref{skyrmionfig}, where a non-trivial winding number $w = \pm 1$ emerges when the vector $\bs{n}(\bs k)$ has covered the whole Bloch sphere once. We remind the reader  that this full covering of the Bloch sphere is achieved when the vector field $\bs{n}(\bs k)$ reaches the North ($\bs n=+ \bs{1}_z$) and South ($\bs n = -\bs{1}_z$) poles  at the two inequivalent Dirac points $\bs K_{\pm}$.  In Fig. \ref{skyrmionfig}, we observe a radical change in the behavior of $\bs n (\bs K_{\pm})$ as $p_y$ is varied. For example, for $p_y=K_y$ (Fig. \ref{skyrmionfig} (a)), the vector field $\bs n (\bs k)$ visits the North pole twice (i.e. at $\bs K_+$ \emph{and} $\bs K_-$) but never the South pole ($w=0$), while for $p_y= 4 K_y$ (Fig. \ref{skyrmionfig} (c)) the vector visits the entire Bloch sphere once ($w=1$). Between these two topologically different configurations, a direct gap closing occurs at $\bs k = \bs K_{-}$ for $p_y= 2 K_y$ (cf. Fig. \ref{skyrmionfig} (b)), a singular situation where the gapless phase is equivalent to a standard semi-metal \cite{Wallace1947,CastroNeto2009}.  We note that transitions $w=0 \leftrightarrow w =\pm 1$ require a single gap closing point (cf. Fig. \ref{skyrmionfig} (b)), while transitions $w=+1 \leftrightarrow w =- 1$ involve two gap-closing points (cf. Fig. \ref{skyrmionfig} (d)). Therefore, in agreement with the equivalence \eqref{chernwind2}, we observe that the topological phase transitions between different topological semi metals are of the same nature as the transitions between different Chern insulators, in the sense that both phenomena occur through direct gap closing (driven here by the control parameter $p_y$).  \\

In summary, we conclude that the laser-coupled honeycomb lattice and the ToF method of Ref. \cite{Alba2011} offers the possibility to explore the topological order of topological semi-metals, which survive in the absence of a band gap. However, we remind the reader that this detection scheme relies on the evaluation of the winding number $w_{\text{ToF}}$, through a ToF measurement of the vector field $\bs n (\bs k)$, which only converges towards the quantized value $w$ for a complete filling of the lowest energy band $E_- (\bs k)$. Thus, the experimental detection of  topological semi-metals would constitute a subtle task, in the sense that the Fermi energy should be finely tuned in order to maximize the filling of the lowest band (Fig. \ref{gaptopology} (d)). Let us stress that the winding number can only take three possible values, $w=0, \pm1$. Therefore, an experimental plateau $w_{\text{ToF}}(p_y) \sim \pm 1$, stemming from a slightly incomplete filling of the band $E_- (\bs k)$ and from finite size effects, would already provide an acceptable witness of non-trivial topological order. \\

We end this Section by observing that the transitions between topologically different semi-metals are driven by the laser recoil momentum $\brec$, and therefore, this interesting effect cannot be captured by the original Haldane model.

\section{Conclusion}\label{conclusions}

In this work, we explored the rich properties of the laser-coupled honeycomb lattice, which is described by the Hamiltonian \eqref{hamtwo}-\eqref{sz}. We demonstrated the existence of robust Chern insulators in this system, which can be reached in experimentally accessible regions of the large parameter space. In particular, we showed that the possibility of producing such non-trivial phases highly depends on the laser-coupling, through the orientation of the momentum transfer $\brec$ and the effective (laser-induced) tunneling amplitude $t$. We showed that it is important to finely tune the  ratios $t_A /t_B$ and $t_{A,B}/t$ in order to open large and robust topological bulk gaps of the order $\Delta \sim 2 t$. We also discussed the role of the staggered potential $\varepsilon$, which is shown to be crucial in the fully symmetric regime $t=t_{A,B}$, and which could also be used to drive transitions between topological phases of different nature. Importantly, we addressed the question of detectability in the context of the quest for robust Chern insulators, and we stressed the importance of identifying regimes corresponding to large bulk gaps. We showed that an experimental measure of the topological winding number  \eqref{qiformula}-\eqref{dis2}, e.g. through ToF measurement \cite{Alba2011}, yields a strong signature for two types of topological phases: the Chern insulating phase and the topological semi-metal (a semi-metal characterized by a non-trivial winding number). Importantly, we showed that the detection of the topological semi-metal would require a delicate tuning of the Fermi energy, which privileges the search for Chern insulating phases from an experimental point of view. \\

The Chern insulator could alternatively be detected through the identification of chiral edge states, which are protected by the topological gap. A clear signature could be obtained, for example, using the shelving method described in Ref. \cite{Goldman2012}. From the spectra presented in Figs. \ref{figa7}(a),(c) and Fig. \ref{skyrmionfig}, we find that these topological edge states remain robust  in the topological semi-metallic phase: the edge states can only disappear from the bulk gap through direct band-touching processes at the Dirac points. However, the experimental identification of these robust edge states for the semi-metallic regime, e.g. using the shelving method, remains an open question to be explored. \\

Let us finally end this work by mentioning the fact that this system could be directly extended to reproduce the spinful Kane-Mele model for $\mathbb{Z}_2$ topological insulators \cite{Kane:2005} (see also its generalizations \cite{Goldman:2012,Beugeling:2012,Shevtsov:2012,Goldman:2011,Guo:2009,Lan:2012}). In this case, each triangular sublattice should trap atoms in two internal atomic states (Zeeman sublevels), yielding a ``spin"-1/2 structure. These atoms should then be coupled independently by lasers in such a way that the tunneling operators, which are $2 \times 2$ matrices acting between NN sites $n_A$ and $m_B$, have the form $$U (n_A, m_B)= \exp (i \sigma_Z \brec \cdot (\bs r_{n_A} + \bs r_{m_B})/2),$$ where $\sigma_Z$ acts on the ``spins" and $U (n_A, m_B)=U^{\dagger} (m_B,n_A)$. In this spinful honeycomb lattice configuration, non-trivial $\mathbb{Z}_2$ topological phases featuring helical edge states \cite{Kane:2005,Bernevig:2006,Wu:2006}, should be reached in the non-trivial regions identified in Section \ref{topologicalphases}. Thus, the versatile laser-coupled honeycomb lattice is well suited for the exploration of two-dimensional topological phases with cold atoms \cite{Goldman:2010,Stanescu:2010,Beri:2011,Mazza:2012,Goldman2012,Buchhold:2012,Cocks:2012}.

\paragraph*{Acknowledgments}
We thank the Lithuanian Research Council, F.R.S-F.N.R.S (Belgium), DARPA (Optical lattice emulator project), the Emergences program (Ville de Paris and UPMC), the Carnegie Trust for the Universities of Scotland, EPSRC,  and ERC (Manybo Starting Grant) for financial support. I.B.S acknowledges the financial support the NSF through the Physics Frontier Center at JQI, and the ARO with funds from both the Atomtronics MURI and the DARPA OLE Program. The authors thank J. Dalibard, J. Beugnon, S. Nascimb\`ene and N. Bray-Ali for stimulating discussions.

\section*{References}



\providecommand{\newblock}{}



\appendix

\section{Analytical calculation of the Chern number}
\label{analytical}

In this Appendix, we provide a more detailed calculation of the Chern number \eqref{chernone}, which further highlights the role played by the singularities at the Dirac points $\bs{K}_{\pm}$. First, we express the Chern number as 
\begin{align}
  \nu&= \frac{1}{2 \pi} \int_{\mathbb{T}^2} \bs 1_z \cdot (\nabla _{\bs k} \times \bs A (\bs k)  ) \, \txt{d}^2 \bs{k},  \nonumber \\
  &= \frac{1}{2 \pi} \int_{\mathbb{T}^2} \mathcal{F}  , \label{chernappendix} 
\end{align}
where the Berry's curvature  $\mathcal{F}=F_{xy} \, \txt{d}k^x \wedge \text{d} k^y$ is a two-form associated with the Berry's connection $\mathcal{A}= A_{\mu} \, \txt{d}k^{\mu}=  i  \langle u_{(-)} \vert  \nabla_{\mu} \vert u_{(-)} \rangle \, \txt{d}k^{\mu}$ through the exterior derivative $\mathcal{F}=\text{d} \mathcal{A}$, with
\begin{align}
F_{xy} (\bs k)&= \partial_{k_x} A_y (\bs k) - \partial_{k_y} A_x (\bs k)   . 
\end{align}
Note that contributions due to any (gauge-dependent) singularities of $\bs A(\bs k)$ should be excluded in the first-line of Eq. \eqref{chernappendix}.

Considering the gauge in which the lowest eigenstate of the Hamiltonian \eqref{hamcoupling} is given by
\be
\qquad \vert u_{(-)} \rangle =\begin{pmatrix}
-e^{-i \phi} \sin (\theta /2) \\
\cos (\theta /2)
\end{pmatrix},
\ee
we find
\begin{align}
A_{\mu} (\bs k)&=\frac{1}{2} (1- \cos \theta)  \nabla_{k_{\mu}} \phi ,\label{gaugeappendix} \\
F_{xy} (\bs k)&=\frac{1}{2} \sin \theta \bigl ( \partial_{k_x} \theta \, \partial_{k_y} \phi - \partial_{k_y} \theta \,\partial_{k_x} \phi    \bigr ). \nonumber
\end{align}
At this point, let us note that the Berry's curvature $\mathcal{F}$ is a gauge invariant quantity, which remains well defined over the entire FBZ. In contrast, the Berry's connection $\mathcal{A}$ depends on the gauge and can potentially possess singularities within the FBZ. In the present gauge, the singularities correspond to $\cos \theta (\bs k) = -1$, which can only happen at a Dirac point $\bs{K}_D$, under the condition that $d_z(\bs{K}_D) <0$   (cf. also main text). We stress that such singularities, if present, could either take place at one or at two inequivalent Dirac points $\bs{K}_D=\bs K_{\pm}$, depending on the model parameters $(t_{A,B}$, $\varepsilon$, $\bs{p})$ that determine the specific values of $d_z(\bs{K}_{\pm})$. In the following, we will show that it is the \emph{number of singularity points} that determines the non-triviality of the Chern number in Eq. \eqref{chernappendix}. To do so, let us consider the following situations:

\subsubsection*{Absence of singularities}
\label{nosing}

When the Berry's connection is regular over the entire FBZ, the Berry's curvature $\mathcal{F}=\text{d} \mathcal{A}$ is an \emph{exact differential form}. In this case, the Chern number \eqref{chernappendix} is given by the integral over a closed manifold (i.e. the two-torus $\mathbb{T}^2$) of an exact differential form,
\begin{align}
  \nu= \frac{1}{2 \pi} \int_{\mathbb{T}^2} \mathcal{F} = \frac{1}{2 \pi} \int_{\mathbb{T}^2} \text{d} \mathcal{A} = 0 , 
\end{align}
which is trivial from Stokes theorem. In particular, when $d_z(\bs{K}_{\pm}) > 0$ at the two inequivalent Dirac points, the Berry's connection  \eqref{gaugeappendix} remains regular over the entire FBZ and the Chern number necessarily vanishes. \\

\begin{figure}[!]
\centering
\includegraphics[width=0.7\columnwidth]{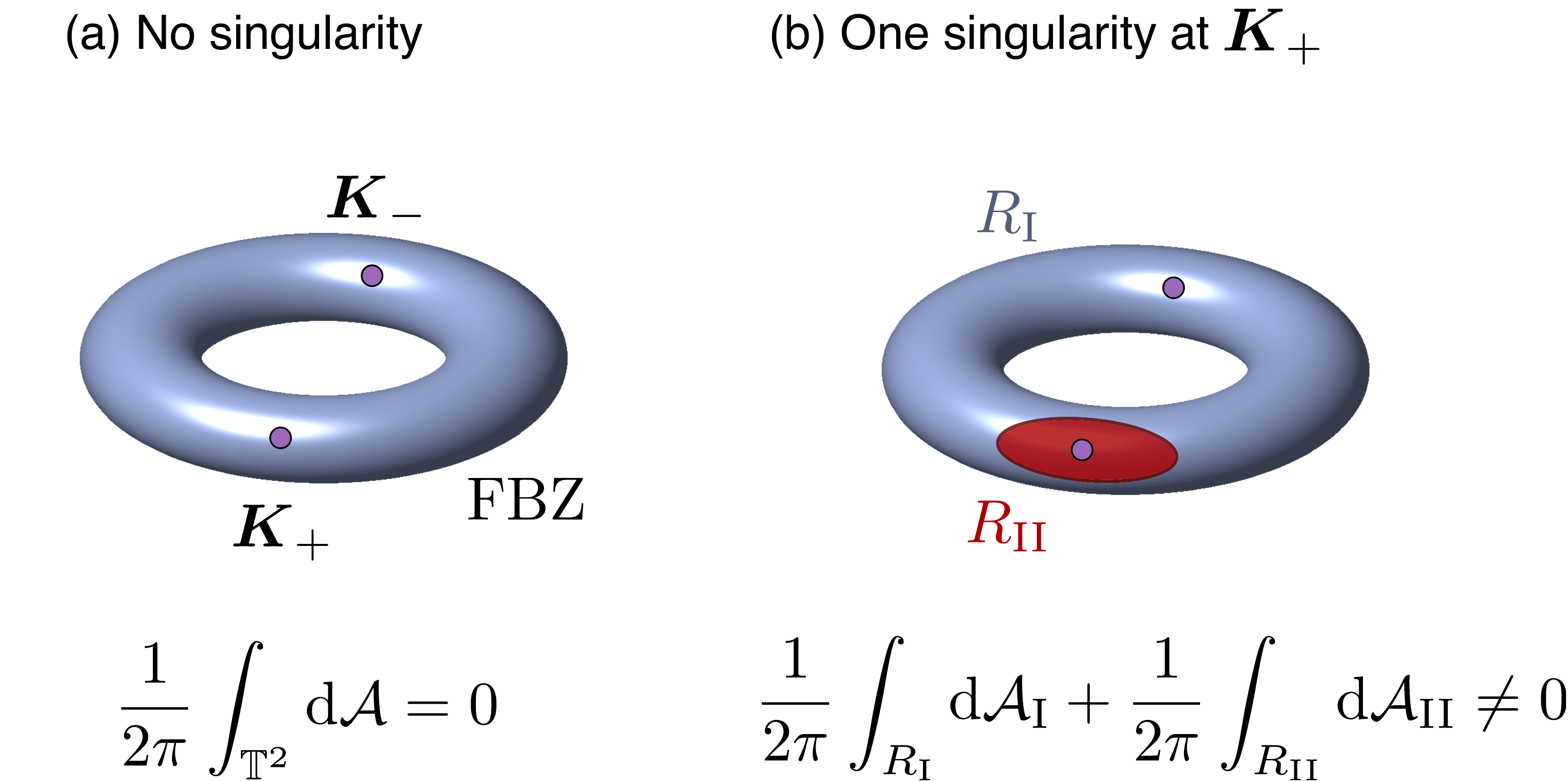}
\caption{The first Brillouin zone (FBZ) and the patchwork configuration. (a) When the Berry's connection  \eqref{gaugeappendix} is regular over the entire FBZ, the Chern number \eqref{chernappendix} vanishes, by direct application of Stokes theorem to the edgeless torus. (b) When the Berry's connection  \eqref{gaugeappendix} is singular at a unique Dirac point (e.g. $\bs K_{+}$), the Chern number is non-trivial. In this case, one defines a local Berry's connection $\mathcal{A}=\{ \mathcal{A}_{\text I} , \mathcal{A}_{\text{II}} \} $, which is regular within the two corresponding regions $R_{\text{I}}$ and $R_{\text{II}}$. Here, the common boundary between these two regions is $\partial R = \gamma_+$.} \label{topologyfig}
\end{figure}

\subsubsection*{One singularity at a unique Dirac point}

Now suppose that the Berry's connection $\mathcal{A}$ in Eq. \eqref{gaugeappendix} is singular at \emph{one} Dirac point, say $\bs{K}_D=\bs K_{+}$. This situation occurs when $d_z(\bs{K}_{+}) < 0$ and $d_z(\bs{K}_{-})>0$, which we now consider to be the case in this paragraph. In the presence of such a singularity, the Berry's curvature $\mathcal{F}$ is no longer an exact differential form, and Stokes theorem cannot be applied globally over the torus. In order to compute the integral \eqref{chernappendix}, we partition the FBZ into two complementary regions, $R_{\text I}$ and $R_{\text{II}}$, whose common boundary $\partial R$ is chosen to be a loop $\gamma_{+}$ encircling $\bs K_+$. Here, we define the region $R_{\text{II}}$ as the one that contains the Dirac point $\bs{K}_{+}$ at which the singularity takes place (cf. Fig. \ref{topologyfig} (b)). Then, we define specific gauges within each region \cite{Wu1975,Kohmoto:1985,Hatsugai1993}, 
\be
\vert u_{(-)} \rangle_{\text{I}} =\begin{pmatrix}
-e^{-i \phi} \sin (\theta /2) \\
\cos (\theta /2)
\end{pmatrix}, \, \vert u_{(-)} \rangle_{\text{II}} =\begin{pmatrix}
- \sin (\theta /2) \\
e^{i \phi} \cos (\theta /2)
\end{pmatrix}.
\ee
In this patchwork configuration, the Berry's connection $\mathcal{A}=\{ \mathcal{A}_{\text I} , \mathcal{A}_{\text{II}} \} $ is a locally-defined quantity, which is now given by
\begin{align}
\bs{A}_{\text I} (\bs k)&=\frac{1}{2} (1- \cos \theta)  \bs{\nabla}_{\bs k} \phi ,\label{gaugeappendix2} \\
\bs{A}_{\text{II}} (\bs k)&= - \frac{1}{2} (1 + \cos \theta)  \bs{\nabla}_{\bs k} \phi ,\label{gaugeappendix3}
\end{align}
inside the regions $R_{\text{I}}$ and $R_{\text{II}}$, respectively. We note that the gauge structures of the two individual regions are connected at the frontier $\partial R=\gamma_{+}$ through the gauge transformation
\begin{align}
&\vert u_{(-)} \rangle_{\text{II}}= e^{i \phi(\bs k)} \vert u_{(-)} \rangle_{\text{I}} \\
&\bs{A}_{\text{II}} (\bs k)=\bs{A}_{\text I} (\bs k) - \bs{\nabla}_{\bs k} \phi.
\end{align}
Furthermore, we note that the Berry's connection $\bs{A}_{\text{II}} (\bs k)$ is now regular at the Dirac point $\bs{K}_{+}$, where $d_z(\bs{K}_{+}) < 0$.   Therefore, the locally-defined Berry's connection $\mathcal{A}=\{ \mathcal{A}_{\text I} , \mathcal{A}_{\text{II}} \}$ is regular over the entire FBZ, and the integral \eqref{chernappendix} can now be computed by applying Stokes theorem to the two different regions \cite{Kohmoto:1985,Hatsugai1993}
\begin{align}
  \nu&=\frac{1}{2 \pi} \int_{\mathbb{T}^2} \mathcal{F} = \frac{1}{2 \pi} \int_{R_{\text{I}}} \text{d} \mathcal{A}_{\text I} +  \frac{1}{2 \pi} \int_{R_{\text{II}}} \text{d} \mathcal{A}_{\text{II}} ,  \nonumber \\
  &=\frac{1}{2 \pi} \oint_{\partial R} \bigl ( \bs{A}_{\text{II}} - \bs{A}_{\text{I}} \bigr ) \cdot \text{d} \bs{k} \nonumber \\
&  \nonumber \\
&= - \frac{1}{2 \pi} \oint_{\gamma_{+}} \bs{\nabla}_{\bs k} \phi (\bs k)  \cdot \text{d} \bs{k}= - v_{+} /2 \pi= - 1 . \label{pos}
\end{align}
Therefore, when the singularity only takes place at the Dirac point $\bs{K}_{+}$, the Chern number is non-trivial and its value is directly related to the vorticity $v_+$ associated with this Dirac point \cite{Hatsugai1993}. \\

In the opposite situation, where the singularity only takes place at the other Dirac point $\bs K_D=\bs{K}_{-}$, namely when $d_z(\bs{K}_{+}) > 0$ and $d_z(\bs{K}_{-}) < 0$,  a similar calculation  (with $\partial R = \gamma_-$) yields
\be
\nu = - v_- /2\pi = + 1 . \label{neg}
\ee

\subsubsection*{Singularities at both Dirac points}

When the Berry's connection \eqref{gaugeappendix} is singular at both Dirac points, namely when $d_z(\bs{K}_{\pm}) < 0$, the Chern number \eqref{chernappendix} is necessarily trivial. Indeed, the gauge transformation
\be
\vert u_{(-)} \rangle \rightarrow \vert \tilde{u}_{(-)} \rangle = e^{i \phi(\bs k)} \vert u_{(-)} \rangle ,
\ee
simultaneously removes the singularities at both Dirac points. In this case, Stokes theorem can be applied globally over the entire FBZ, leading to a zero Chern number (cf. the case with no singularity).  \\

\subsubsection*{Synthesis}

From the results presented above, we conclude that the Chern number $\nu$ characterizing the topological order of our system can only take non-trivial values $\nu=\pm 1$ when the Berry's connection \eqref{gaugeappendix} features a unique singularity inside the FBZ. Therefore, this non-trivial regime is reached when the function $d_z (\bs k)$ has \emph{opposite signs} at the two inequivalent Dirac points $\bs {K_{\pm}}$. Then, from Eqs. \eqref{pos}-\eqref{neg}, we finally obtain the result
\begin{align}
  \nu=\frac{1}{2} \biggl( \text{sign} \bigl (d_z (\bs K_{+}) \bigr) - \text{sign} \bigl (d_z (\bs K_{-}) \bigr )   \biggr ),
\end{align}
already announced in Eq. \eqref{chern2}.

\section{Calculation of topological edge states in finite geometries }\label{app:edge}

It is a standard procedure to determine the edge state structure by considering a semi-infinite system, namely using a cylindrical geometry in which periodic boundary conditions have only been applied to one spatial direction. Here, we assume that the system is closed along the $y$ direction only, and we write the single-particle eigenfunctions as $\psi_{A,B} (\bs r)= \exp (i k_y y) \, u_{A,B} (\bs r)$, where $u_{A,B} (\bs r + \bs a_3)=u_{A,B} (\bs r)$. Setting $\Psi_n= (u_{A} (\bs r_{n_A}), u_{B} (\bs r_{n_A} - \bs \delta_2))$, where the index $n$ labels the sites along the open direction (chosen along $x$ here), we obtain the Harper-like equation
\begin{align}
&E  \Psi_n
  = \mathcal{D}  \Psi _n\!+\!  \mathcal{R} \Psi_{n+1}\! +\! \mathcal{R}^{\dagger} \Psi_{n-1} , \notag \\
& \mathcal{D}
=\begin{pmatrix}
- \varepsilon - 2 t_A \cos \bigl ( \brec \cdot \bs a_3/2 - k_y (a_3)_y \bigr ) & -t \bigl ( e^{-i k_y (\delta_2)_y} + e^{-i k_y (\delta_1)_y} \bigr ) \\
-t \bigl ( e^{i k_y (\delta_2)_y} + e^{i k_y (\delta_1)_y} \bigr ) &  \varepsilon - 2 t_B \cos \bigl ( \brec \cdot \bs a_3 /2 + k_y (a_3)_y  \bigr ) 
\end{pmatrix}, \notag \\
& \mathcal{R}
=\begin{pmatrix}
- t_A \biggl ( e^{-i  \brec \cdot \bs a_1/2} e^{i k_y (a_1)_y} +  e^{-i  \brec \cdot \bs a_2/2} e^{i k_y (a_2)_y}  \biggr ) &  -t e^{-i k_y (\delta_3)_y} \\
0 & - t_B \biggl ( e^{i  \brec \cdot \bs a_1/2} e^{i k_y (a_1)_y} +  e^{i  \brec \cdot \bs a_2/2} e^{i k_y (a_2)_y}  \biggr )
\end{pmatrix}.\label{eq:cylindre}
\end{align}
The energy spectrum $E=E(k_y)$, describing the bulk but also the edge states, can be obtained by solving the corresponding $2 L \times 2 L$ Hamiltonian matrix numerically, where $n=1, \dots, L$. We stress that the chiral edge states,  identified with this method, could lead to clear signatures in an optical-lattice setup, even in the presence of an external confining trap \cite{Stanescu:2010,Goldman2012,Buchhold:2012}. 

\section{The winding number and finite size effects}\label{app:finite}

\begin{figure}
\centering
\includegraphics[width=1\columnwidth]{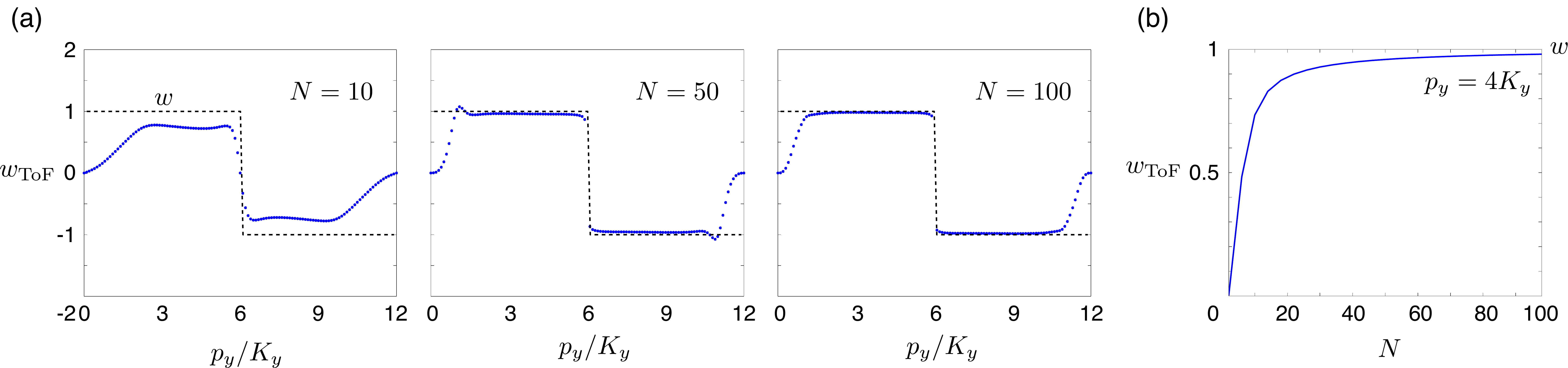}
\caption{(a)The winding number $w_{\text{ToF}}$ as a function of $p_y$ for $t_A=t_B=0.3 t$, $\varepsilon=0$. The winding number $w_{\text{ToF}}$ has been computed through Eq. \eqref{dis2}, using a $N \times N$ lattice. Here, the Fermi energy is set such that the lowest energy band $E_- (\bs k)$ remains perfectly filled. (b) The winding number $w_{\text{ToF}}$ as a function of the lattice length $N$, for $p_y=4 K_y$, $t_A=t_B=0.3 t$ and $\varepsilon=0$. } \label{finitefig}
\end{figure}

In this Appendix, we analyze the finite size effects that arise when the winding number $w_{\text{ToF}}$ is evaluated through the discrete sum \eqref{dis2}. The results are presented in Fig. \ref{finitefig}, in the ideal case where the Fermi energy is chosen such that the lowest energy band remains perfectly filled. The winding number $w_{\text{ToF}}$ is computed for different lattice sizes $N \times N$, with $N=10,50,100$. The convergence of the winding number $w_{\text{ToF}}$ towards the quantized value $w=+1$ is shown in Fig. \ref{finitefig}(b).  

\section{The Hall conductivity and the winding number}\label{app:hall}

The Hall conductivity is given by the Kubo formula \cite{Thouless1982}
\begin{align}\label{sigmaxy}
\sigma_{H}&= \frac{e^2}{\hbar} \frac{i}{V} \sum_{E_{\alpha} < E_{\text{F}}}   \langle \partial_{k_x} u_{\alpha}(\bs k) \vert \partial_{k_y} u_{\alpha} (\bs k) \rangle
  - (k_x \leftrightarrow k_y), \nonumber \\
  &= \frac{e^2}{h} \frac{i}{2 \pi} \Delta k_x \Delta k_y  \sum_{E_{\alpha} < E_{\text{F}}}   \langle \partial_{k_x} u_{\alpha}(\bs k) \vert \partial_{k_y} u_{\alpha} (\bs k) \rangle
  - (k_x \leftrightarrow k_y),
\end{align}
where $V$ is the volume and where the sum $\sum_{\alpha}$ takes into account the contribution of all the occupied states $ \vert u_{\alpha} \rangle= \vert u_{\pm} (\bs k) \rangle$ with energy $E_{\alpha}=E_{\pm}(\bs k)<E_{\text{F}}$. 
In our two-band system, the sum in Eq.\eqref{sigmaxy} can be decomposed into two parts 
\begin{align}
&\sum_{\alpha} F_{xy}^{\alpha} (\bs k)= \Biggl ( \sum_{\mathcal{K}_{(-)}} - \sum_{\mathcal{K}_{(+)}} \Biggr ) F_{xy}^{(-)} (\bs k), \\
&\mathcal{K}_{(\pm)}=\{\bs k: E^{(\pm)}(\bs k_{\pm}) < E_F\}, \nonumber
\end{align}
where $F_{xy}^{(\pm)} (\bs k)=\partial_{k_x} A_y^{(\pm)} (\bs k) - \partial_{k_y} A_x^{(\pm)} (\bs k)$ is the Berry's curvature associated with the state $ \vert u_{(\pm)} (\bs k) \rangle$,  and where we used the fact that $F_{xy}^{(-)} (\bs k) = - F_{xy}^{(+)} (\bs k)$. 
When the first band $E_{-} (\bs k)$ is totally filled ($\mathcal{K}_{(-)}=\text{FBZ}$, $\mathcal{K}_{(+)}=\emptyset$), namely when $E_{\text{F}}$ lies in a spectral gap, we find the usual TKNN relation (or Chern number $\nu$) \cite{Thouless1982}
\be
\sigma_{xy}= \frac{e^2}{h} \frac{1}{2 \pi} \int_{\mathbb{T}^2} F_{xy}^{(-)} (\bs k) \, \text{d} \bs k = \frac{e^2}{h} \nu,
\ee
in the limit $\Delta k_{x,y} \rightarrow 0$. When the Fermi energy is not located in a bulk gap, the Hall conductivity must be computed using the more general expression
\be
\sigma_{xy}= \frac{e^2}{h} \frac{1}{2 \pi}  \Biggl ( \int_{\mathcal{K}_{(-)}}\text{d} \bs k - \int_{\mathcal{K}_{(+)}}\text{d} \bs k \Biggr ) F_{xy}^{(-)} (\bs k) , \label{equdisc}
\ee
which takes into account the fact that both bands $E_{\pm} (\bs k)$ could be partially filled. \\

Next, we note that the Berry's curvature $F_{xy}=F_{xy}^{(-)}$, given in Eq. \eqref{gaugeappendix}, is equal to the Pontryagin form
\be
F_{xy} (\bs k)=\frac{1}{2} \sin \theta \bigl ( \partial_{k_x} \theta \, \partial_{k_y} \phi - \partial_{k_y} \theta \,\partial_{k_x} \phi    \bigr )= \frac{1}{2} \bs n \cdot \biggl ( \partial_{k_x} \bs n \times  \partial_{k_y} \bs n  \biggr) ,\label{pontberry}
\ee
where the vector field $\bs n (\bs k)=\bs d (\bs k) / d (\bs k)$ is defined in Eq. \eqref{sz}. Therefore, we can write the Hall conductivity in terms of the vector field $\bs n (\bs k)$
\be
\sigma_{xy}= \frac{e^2}{h} \frac{1}{4 \pi}  \Biggl ( \int_{\mathcal{K}_{(-)}}\text{d} \bs k - \int_{\mathcal{K}_{(+)}} \text{d} \bs k \Biggr )  \bs n \cdot \biggl ( \partial_{k_x} \bs n \times  \partial_{k_y} \bs n  \biggr) .  \label{equdisc2}
\ee
Discretizing the Pontryagin form 
\begin{align}
&\bs n \cdot \biggl ( \partial_{k_x} \bs n \times  \partial_{k_y} \bs n  \biggr)=\Delta k_x^{-1} \Delta k_y^{-1} \sum_{\nu \ne \mu \ne \lambda}  n_{\mu} (\bs k) \biggl (  n_{\nu} (\bs k + \bs e_x) n_{\lambda} (\bs k + \bs e_y) - n_{\nu} (\bs k +  \bs e_y) n_{\lambda} (\bs k +  \bs e_x) \nonumber \\
 &\qquad \qquad + n_{\nu} (\bs k + \bs e_y) n_{\lambda} (\bs k)  - n_{\nu} (\bs k +  \bs e_x) n_{\lambda} (\bs k) + n_{\nu} (\bs k) n_{\lambda} (\bs k +  \bs e_x)  - n_{\nu} (\bs k) n_{\lambda} (\bs k +  \bs e_y) ,
\end{align}
with $\nu, \mu, \lambda = x, y,z$, and writing the integrals in \eqref{equdisc2} as sums, leads to the equivalence between the Hall conductivity in Eq. \eqref{equdisc} and the ToF winding number \eqref{dis2}: $\sigma_H = (e^2/h) \, w_{\text{ToF}}$. When the Fermi energy is in a gap, Eq. \eqref{pontberry} also shows the equality between the Chern number $\nu$ and the winding number $w$ (cf. main text).

%

%
%
%
%

\end{document}